\documentclass[aps,reprint,preprintnumbers,twocolumn,floatfix,superscriptaddress,prx,showpacs]{revtex4-2}
\pdfoutput=1
\usepackage{physics}
\usepackage{natbib}
\usepackage{amsmath,amssymb}
\usepackage{graphicx}
\usepackage[dvipsnames]{xcolor}
\usepackage[caption=false]{subfig}
\usepackage{float}
\usepackage[pdftex]{hyperref}
\usepackage[T1]{fontenc}

\begin{document}
\title{Hybrid Quantum Repeaters with Ensemble-based Quantum Memories and Single-spin Photon Transducers}
\author{Fenglei Gu}
\affiliation{QuTech, Delft University of Technology, Lorentzweg 1, 2628 CJ Delft, The Netherlands}
\author{Shankar G Menon}
\affiliation{Pritzker School of Molecular Engineering, University of Chicago, Chicago, IL, 60637, USA}
\author{David Maier}
\affiliation{QuTech, Delft University of Technology, Lorentzweg 1, 2628 CJ Delft, The Netherlands}
\author{Antariksha Das}
\affiliation{QuTech, Delft University of Technology, Lorentzweg 1, 2628 CJ Delft, The Netherlands}
\author{Tanmoy Chakraborty}
\affiliation{QuTech, Delft University of Technology, Lorentzweg 1, 2628 CJ Delft, The Netherlands}
\author{Wolfgang Tittel}
\affiliation{QuTech, Delft University of Technology, Lorentzweg 1, 2628 CJ Delft, The Netherlands}
\affiliation{Department of Applied Physics, University of Geneva, 1211 Geneva 4, Switzerland}
\affiliation{Constructor University, 28759 Bremen, Germany}
\author{Hannes Bernien}
\affiliation{Pritzker School of Molecular Engineering, University of Chicago, Chicago, IL, 60637, USA}
\author{Johannes Borregaard}
\affiliation{QuTech, Delft University of Technology, Lorentzweg 1, 2628 CJ Delft, The Netherlands}
\affiliation{Department of Physics, Harvard University, Cambridge, Massachusetts 02138, USA}

\begin{abstract}
Reliable quantum communication over hundreds of kilometers is a daunting yet necessary requirement for a quantum internet. To overcome photon loss, the deployment of quantum repeater stations between distant network nodes is necessary. A plethora of different quantum hardware is being developed for this purpose, each platform with its own opportunities and challenges. Here, we propose to combine two promising hardware platforms in a hybrid quantum repeater architecture to lower the cost and boost the performance of long-distance quantum communication. We outline how ensemble-based quantum memories combined with single-spin photon transducers, which can transfer quantum information between a photon and a single spin, can facilitate massive multiplexing, efficient photon generation, and quantum logic for amplifying communication rates. As a specific example, we describe how a single Rubidium (Rb) atom coupled to nanophotonic resonators can function as a high-rate, telecom-visible entangled photon source with the visible photon being compatible with storage in a Thulium-doped crystal memory (Tm-memory) and the telecom photon being compatible with low loss fiber propagation. We experimentally verify that Tm and Rb transitions are in resonance with each other. Our analysis shows that by employing up to 9 repeater stations, each equipped with two Tm-memories capable of holding up to 625 storage modes, along with four single Rb atoms, one can reach a quantum communication rate of about 10 secret bits per second across distances of up to 1000 km.
\end{abstract}

\maketitle

\section{Introduction}

The ability to transmit quantum information reliably between distant parties is a prerequisite for any useful application of a quantum internet~\cite{Kimble2008,Wehner2018}. The primary challenge to achieve this is the exponential attenuation of optical signals in fiber-based networks. To overcome this challenge, quantum repeaters have been proposed where the distance is divided into shorter segments over which entanglement can be established in a heralded fashion. Once entanglement has been successfully established over the segments, entanglement swapping can extend the entanglement over the total distance~\cite{Munro2015,azuma2023_RMP}.

Different quantum hardware such as solid-state defect centers~\cite{Hermans2022,knaut2023}, atomic ensembles~\cite{Heller2020,Pu2021,Askarani2021}, trapped ions~\cite{Bock2018,Krutyanskiy2023} and quantum dots~\cite{Neuwirth2021} are currently being developed to enable a functional quantum repeater. There exist numerous theoretical proposals for repeater architectures tailored to the specific features of each hardware~\cite{Sangouard2009,Nemoto2016,Borregaard2020,Sharman2021}.

Quantum repeaters with ensemble-based quantum memories, pioneered by the Duan-Lukin-Cirac-Zoller (DLCZ) protocol~\cite{Duan2001}, have been pursued heavily experimentally due to their technological simplicity and multiplexing capabilities~\cite{Lago2021,Li2021,Businger2022}. However, using ensembles makes it difficult to perform quantum logic on the stored information. Repeater protocols thus resort to probabilistic entanglement swapping schemes based on linear optics, severely limiting the performance. In addition, the probabilistic generation of approximate pairs of entangled photons from either the ensembles themselves or external Spontaneous Parametric Down conversion (SPDC) sources leads to a fundamental trade-off between the rate and fidelity of the communication~\cite{Krovi2016,yoshida2024multiplexed} severely limiting the usefulness of repeaters.

Quantum repeaters based on individual atoms or atom-like defects represent an alternative route~\cite{Hermans2022,knaut2023,Langenfeld2021}. Near-deterministic single-spin photon transducers, which can transfer quantum information between a photon and a single spin, can be achieved by coupling the atomic system to optical resonators enabling efficient single-photon generation~\cite{Morin2021,Knall2022,Raha2020}. The ability to manipulate the hyperfine states of single atoms allows for quantum logic enabling deterministic entanglement swapping and purification techniques~\cite{Kalb2017}. However, repeater protocols based on merely individually trapped atoms make large multiplexing a formidable challenge given current technology.

Here, we propose to combine ensemble-based quantum memories with single-spin photon transducers to enable a near-term hybrid quantum repeater with massive multiplexing, efficient photon generation, and near-deterministic entanglement swapping. In our scheme, a single-spin photon transducer is used for high-rate generation of entangled photon pairs where, for each pair, one photon is to be stored in a multi-mode ensemble-based memory in a repeater node and the other to be transmitted through a fiber to generate distant entanglement over the elementary segment between the nodes. Successfully entangled stored photons are read out from the memories and near-deterministic entanglement swapping can be accomplished with the aid of extra single-spin photon transducers, thereby extending entanglement over neighboring elementary segments.

Furthermore, we outline a specific implementation with cavity-coupled single Rb atoms and Tm-doped crystal memories. We show how a single Rb atom coupled to two nanophotonic cavities with visible and telecom resonance frequencies, respectively, can function as a robust photon-pair source producing entangled visible and telecom photons. The telecom photon can propagate in standard optical fibers with minimum loss~\cite{Tanzilli2005} and we experimentally verify that the visible photon is compatible with the resonance of the Tm-doped crystal. Thus no frequency conversion is required for the repeater. We simulate the performance of the repeater for quantum key distribution and show that rates of tens of secret bits per second over distances of up to 1000 km can be achieved with up to 9 repeater stations, each containing only two ensemble-based memories and four single Rb atoms. Further increase of the rate is possible through additional multiplexing.

The article is organized as follows: Section II introduces the overall protocol and the mechanism of the entanglement swapping. Section III discusses the mechanism by which Rb atoms emit both telecom and visible photons.
Section IV presents an experimental study demonstrating the compatibility between the Rb photon source and quantum memory based on Tm:LiNbO$_3$. In Section V, we show the simulation
results of the repeater chain. Finally, Section VI offers a general discussion and an outlook.

\section{Structure of the Repeater Chain Protocol}

\begin{figure*}
\includegraphics[width=0.95\textwidth]{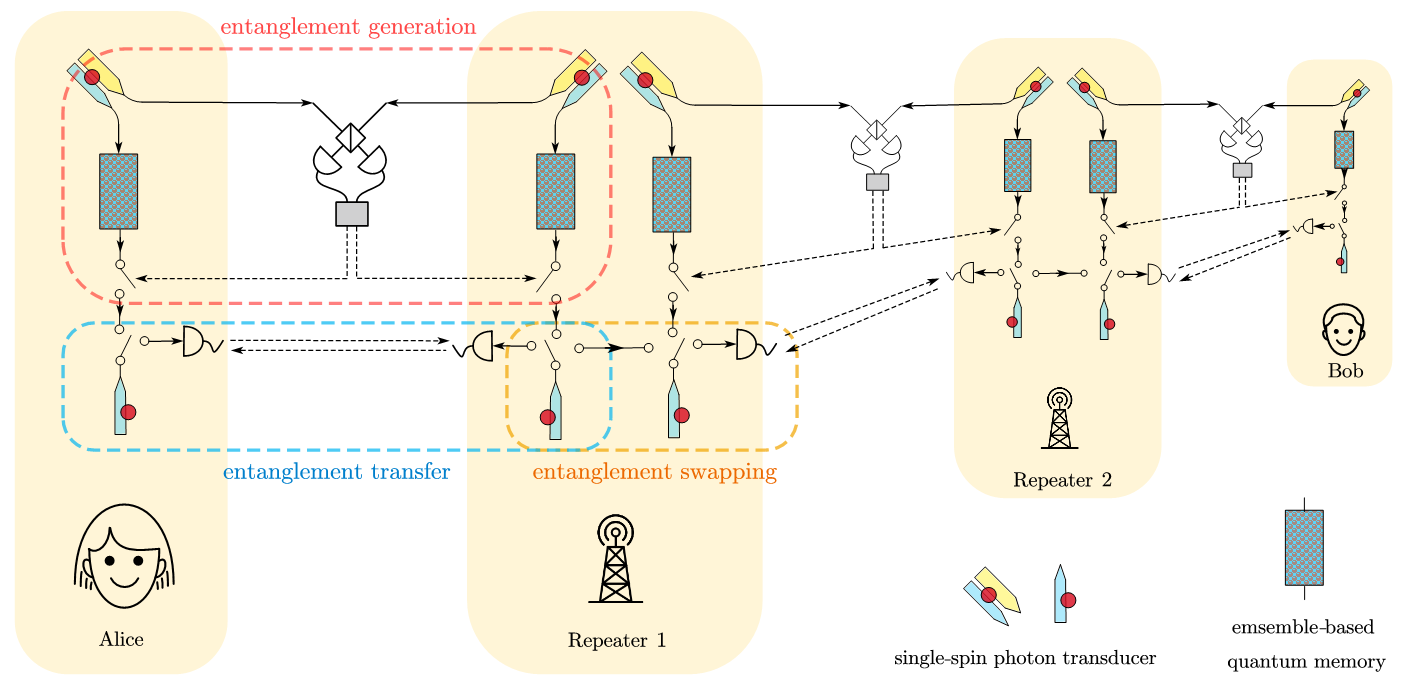}
\centering
\caption{\label{fig_structure}Structure of the quantum repeater chain architecture. The whole system contains two end-nodes, Alice and Bob, connected by a series of repeater nodes. This figure depicts an example with two repeater nodes showcasing the key components involved. The yellow blocks represent the repeater nodes with the local devices depicted within. The protocol involves three main steps: entanglement generation, entanglement transfer, and entanglement swapping, as circled with dashed lines in different colors in the figure. The whole protocol harnesses two kinds of devices as the key components -- the single-spin photon transducers and the ensemble-based quantum memories. The others are common devices including 50-50 optical beam splitters, photon detectors, and optical switches, as sketched with conventional symbols in the figure. The solid-line arrows represent the optical fibers and the dashed-line arrows the classical channels. Both the end nodes and the repeater nodes have symmetric layouts.}
\end{figure*}

The general structure of the quantum repeater, composed of the single-spin photon transducers and ensemble-based quantum memories, is depicted in Fig.~\ref{fig_structure}. The protocol is designed to distribute entanglement between two distant end nodes, Alice and Bob. It divides the total distance between Alice and Bob into multiple segments. In each segment, we use photon transducers to repeatedly emit entangled photon pairs on each side, storing one photon from each pair in the ensemble-based memories and attempting heralded entanglement generation over the segment with the other photons. Upon success, the now entangled photons are retrieved from the memories on each side and mapped into single-atom systems in a heralded fashion. This is followed by entanglement swapping using local Bell measurements to generate entanglement between Alice and Bob. The protocol thus consists of three main steps: entanglement generation, entanglement transfer, and entanglement swapping as circled separately in Fig.~\ref{fig_structure}.

For the initial entanglement generation step, as shown within the dashed red boundary in Fig.~\ref{fig_structure}, two identical single-spin photon transducers are employed to produce entangled photon pairs continuously. For each photon pair, one of the photons is to be stored in the multi-mode ensemble-based quantum memory while the other is sent via optical fiber to the middle station. At the middle station, a linear optics Bell measurement is performed to entangle the photons stored in the quantum memories. While the success of this operation is probabilistic and subjected to photon loss, information about which photon pairs were successfully entangled is sent back from the middle station to the nodes where quantum memories are located. The successfully entangled photons will be read out from the memories for further processing while the failed ones will be discarded.

As illustrated in the area enclosed with the dashed blue line in Fig.~\ref{fig_structure}, the successfully entangled photon pairs read out from the Tm-memories are transferred into the single-spin photon transducer systems.~The entanglement is swapped from the photons to the spins by means of a cavity-mediated controlled phase gate~\cite{Reiserer2014,Kalb2015}, which also heralds successful retrieval of the photons from the Tm-memory. The heralding signals are sent between the nodes to confirm the successful transfer. Once confirmed, the successfully entangled spins will be stored until entanglement swapping can take place with the neighboring link. Unsuccessfully entangled spins will be reset to attempt the next photonic transfer.

Lastly, for the entanglement swapping between neighboring sections, one has to implement a Bell state measurement between the two adjacent spins in one repeater node. In this case, one can let one of the spins re-emit a photon, guide it to interact with the other spin, and finally, detect the photon and both spins in certain bases, as sketched in Fig.~\ref{fig_structure} within the dashed orange rounded rectangle. This operation is deterministic up to local losses. 

Below, we outline a concrete implementation of this protocol with trapped single neutral Rb atoms and Tm-doped crystal quantum memory (Tm-memory) and assess its performance.

\section{Rb entangled-Photon emitter and entanglement generation protocol} \label{section_Rb}

\begin{figure}
    \includegraphics[width=0.47\textwidth]{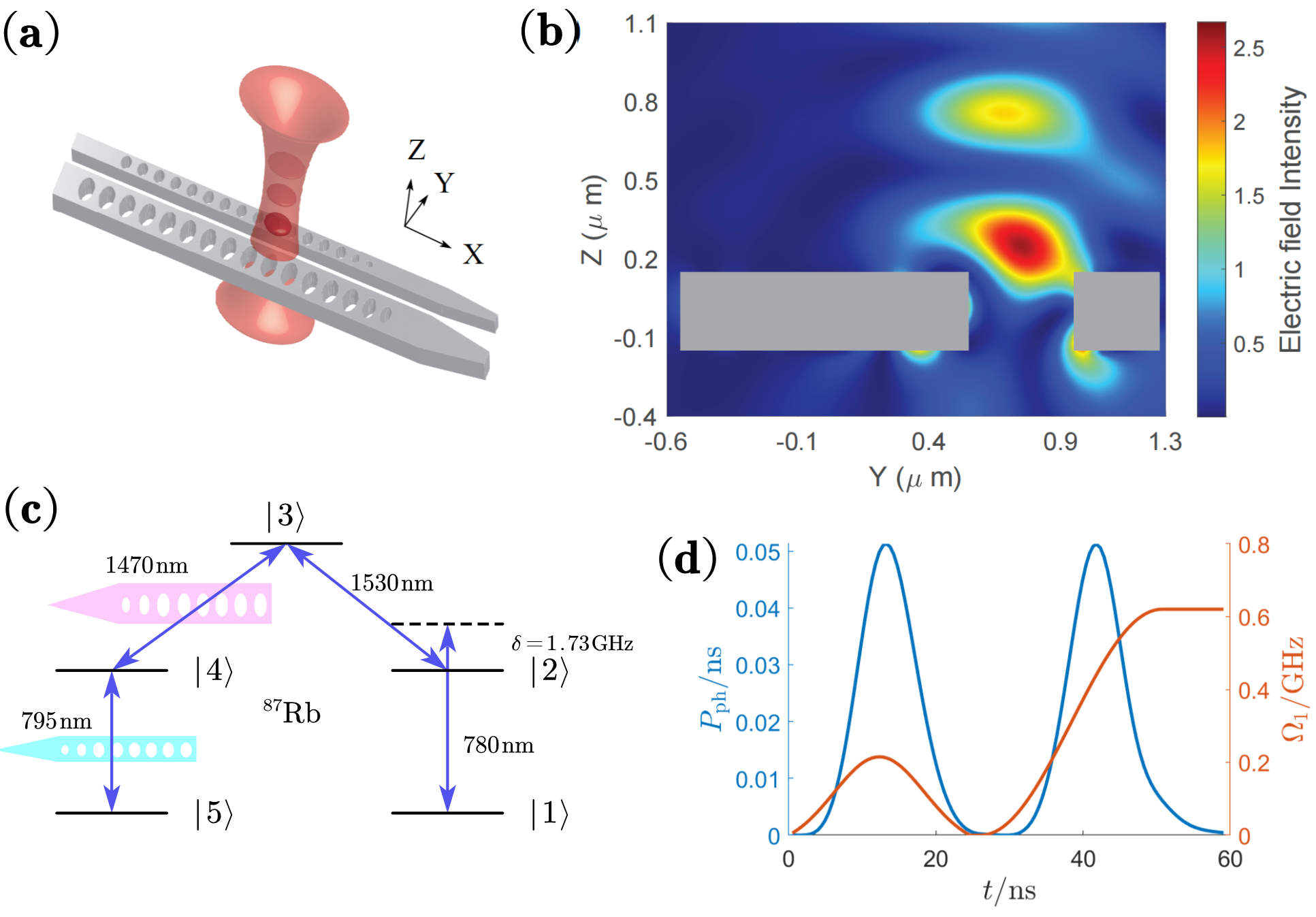}\centering
    \caption{\label{fig_Rb_emitter} Rb emitter design and mechanism. (a) Parallel dual cavity design. The wide and narrow gray strips placed along the X direction represent the nanophotonic cavities resonant with the telecom and visible photons emitted from the Rb atom, respectively. They are of a TE mode telecom wavelength cavity (1470 nm) and a TM mode visible wavelength cavity (795 nm ) both with a refraction index of 2.6 and a thickness of 300 nm. The light-red cone denotes the combination of the incident and reflected trapping lights, forming antinodes of high light intensity marked as dark red. The Rb atom is trapped in the nearest node to the cavities. (b) The slice of the simulated relative electric field intensity of the trapping light of the Z-Y plane centered around the trapping light position. It is normalized to the incident tweezer beam intensity. The two solid gray rectangles indicate the two nanophotonic cavities viewed from the X direction. (c) The intended driving path, from $\ket{1}$ to $\ket{5}$, in the Rb atom. The first two couplings are laser-induced and the later two are cavity-induced. There is a detuning $\delta=1.73$ GHz respecting the $|1\rangle-|2\rangle$ energy difference in the first laser driving. (d) Temporal profile of the driving pulse of the first laser (orange) and emitted telecom and visible photons (blue). The code of this Rb-cavity system simulation can be found in ref~\cite{codeCollection}.}
\end{figure}

The Rb entangled-photon emitter consists of a single neutral Rb atom trapped with optical tweezers and coupled to two nanophotonic cavities, one with resonance wavelength at 1470 nm and the other at 795 nm. There have been previous designs for two-mode cavity coupling of emitters with crossed cavities or using both TE and TM modes in a waveguide~\cite{Zhang2009, Rivoire2011}. It is however challenging for such setups to maintain the required cavity quality factor while increasing the frequency separation of the two modes, which, in our case, corresponds to a wavelength difference of 675 nm. 

Instead, we propose a parallel-cavity arrangement with the Rb atom located on the top of the two cavities, as shown in Fig.~\ref{fig_Rb_emitter}(a). Such an architecture enhances the independence of the two cavities and is feasible by integrating atoms with on-chip nanophotonic cavities~\cite{Menon2023, Kim2019}. However, there are still many issues that need to be considered for choosing suitable system parameters. On the one hand, we need both cavities to be close enough to each other to reflect the tweezer light to form a stable trap for the Rb atoms~\cite{Thompson2013, Samutpraphoot2020}. In addition, placing the cavities closer to each other also provides larger atom-cavity coupling strengths, resulting in Purcell-enhanced emissions into the desired cavity mode. On the other hand, as the cavities of 1470 nm and 795 nm get closer, photonic modes can leak into each other, reducing the achievable quality factors.

We address these challenges by carefully selecting the material with the optimum refractive index, cavity separation, and cavity thickness. In our design, we chose silicon-enriched silicon nitride, SiN, as the cavity material with a refractive index of 2.6. This selection enables better mode confinement compared to stoichiometric ${\rm Si_3N_4}$, allowing for minimal separations between the cavities while maintaining high cavity quality factors. For practical fabrication considerations, both cavities are assumed to have a thickness of 300 nm. We perform simulations with a 1060 nm tweezer and 400 nm separation between the cavities and present the results in Fig.~\ref{fig_Rb_emitter}(b). Our findings demonstrate that this design enables the 1470~nm telecom-photon cavity to achieve a quality factor of $1.4\times10^5$ with an average cooperativity of 34 with the Rb atom, while the 795~nm visible-photon cavity reaches a quality factor of $3.8\times10^5$ with an average cooperativity of 11. More details on the cavity design are provided in Appendix~D.

We will now describe in detail how this Rb-dual-cavity setup can function as a source of entangled photon pairs with the two photons at telecom and visible wavelengths, respectively. To achieve this, we harness five specific electronic orbital states from the $^{87}{\rm Rb}$ atom. They are: $|1\rangle$: $\ensuremath{5^{2}\text{S}_{1/2} |F=2,m_{F}=2}\rangle$, 
$|2\rangle\text{:}\ \ensuremath{5^{2}\text{P}_{3/2} |F=3,m_{F}=3}\rangle$, 
$|3\rangle\text{:}\ 4^{2}\text{D}_{3/2},|F=3,m_{F}=3\rangle$, $|4\rangle\text{:}\ 5^{2}\text{P}_{1/2},|F=2,m_{F}=2\rangle$, 
and $|5\rangle\text{:}\ 5^{2}\text{S}_{1/2},|F=1,m_{F}=1\rangle$. Fig.~\ref{fig_Rb_emitter}(c) shows the coupling and driving of these five levels. In each photon emission cycle, we initialize the Rb atom in state $|1\rangle$ and drive it first to the $|2\rangle$ state with the first laser and then to the $|3\rangle$ state with the second laser. From the $|3\rangle$ state, the Rb atom will decay initially to the $|4\rangle$ state emitting a photon into the telecom cavity mode (1470 nm), followed by another decay to the $|5\rangle$ state emitting a second photon into the visible cavity mode (795 nm). 

The second laser is continuously driving the transition $|2\rangle\leftrightarrow|3\rangle$ at a Rabi frequency of the same order as the cavity couplings, however, the Rabi frequency of the $\ket{1} \leftrightarrow \ket{2}$ drive is chosen to be about an order of magnitude lower. Besides, the pulse of the $\ket{1} \leftrightarrow \ket{2}$ drive is modulated with a pause in the middle of the driving process as shown by the orange curve in Fig.~\ref{fig_Rb_emitter}(d). The paused driving is calibrated such that the telecom and visible photons have equal probabilities of being both generated in the early time-bin (E) and the late time-bin (L), as shown by the blue curve in Fig.~\ref{fig_Rb_emitter}(d). Ideally, this results in an entangled state of the form:
\begin{equation}
    \psi_{{\rm vt}}=\frac{1}{\sqrt{2}}
    \left(|\text{E}\rangle_{\text{v}}|\text{E}\rangle_{\text{t}}+|\text{L}\rangle_{\text{v}}|\text{L}\rangle_{\text{t}}\right),
    \label{eq_ot}
\end{equation}
where the subscripts refer to the visible (v) and telecom (t) photons. We note that the Rb atom is not entangled with the final photonic state and can repeatedly emit states of the form above. We simulated this emission procedure with the comprehensive Hamiltonian and parameters shown in Appendix~A.

In each elementary segment, both end nodes are equipped with Rb emitters. The visible photons generated by the two Rb emitters are directly stored in quantum memories within the nodes while the telecom photons are to be sent via optical fiber to a middle station. At the middle station, a beam splitter is used for erasing the which-way information, and photon detectors after the beam splitter are used to tell in which time-bins the two photons arrive. In this way, with a maximum probability of 50\% for the photon detectors to detect an early as well as a late photon, it ideally projects the two visible photons into a maximally entangled state
\begin{equation}
    \psi_{{\rm vv}}=\frac{1}{\sqrt{2}}
    \left(|\text{E}\rangle_{\text{vl}}|\text{L}\rangle_{\text{vr}} \pm
    |\text{L}\rangle_{\text{vl}}|\text{E}\rangle_{\text{vr}}\right),
    \label{eq_oo}
\end{equation}
where the subscripts "l" and "r" denote the left and right ends and the $\pm$ sign is determined by whether the same detector (+) or different detectors (-) measure a photon in the early and late time bins. This allows one to realize the heralded entanglement generation in each segment in a multiplexed fashion.

In actual implementations, there will be imperfections and noisy processes in the above procedure. To investigate the effect of these, we develop a detailed quantum optical model that includes the many-level structure of the Rb-atoms, faulty laser polarization, finite and fluctuating cavity coupling, and various loss mechanisms. We refer to Appendix~A for details of this model and discuss only some of the key insights here.

Ideally, the first driving laser should drive only the $\sigma_+$ transitions. However, due to the device geometry, good polarization-maintained driving can only be obtained by driving with linear polarization. This results in an equal driving of $\sigma_+$ and $\sigma_-$ transitions. The $\sigma_-$ components leak population to many other hyperfine levels as shown in Appendix~A. This problem can, however, be mitigated by tuning the frequency of the first laser to match a specific resonance of a dressed state resulting from the continuous driving of the excited states from the second laser. The details of this method are presented in Appendix~B.

Besides, the atoms will oscillate in the trapping potential due to finite temperature, which results in fluctuating cavity couplings. While this can in principle be mitigated by cooling the atomic motion~\cite{Thompson_cooling2013, Kaufman2012}, it is important to assess the robustness of our scheme to this. 

Fluctuating cavity couplings mean that photons from different emitters will have varying amplitudes of the photonic time-bins due to the fluctuations. This, in turn, will degrade the quality of the generated entanglement. While this cannot be completely circumvented, we identify certain `sweet spots' of the frequency of the first driving laser where the effect of these fluctuations can be efficiently suppressed. This can be understood from the dressed states of the subspace $\{\ket{2},\ket{3},\ket{4}\}$ coupled by the telecom cavity field and the second laser. By tuning the frequency of the first driving laser, we can target one of the dressed states such that the effect from the fluctuation of $\ket{3}-\ket{4}$ coupling will be counteracted by the effective $\ket{1}-\ket{2}$ coupling. This procedure is described in more detail in Appendix~C.

Adopting the error-suppression techniques described above, our simulation shows that entangled states with an average fidelity of 0.98 with respect to the target state in Eq.~(\ref{eq_oo}) can be achieved. This simulation is based on fluctuating cooperativities of $34\pm 5.0$ for the telecom-photon cavity and $11\pm 2.2$ for the visible-photon cavity. For the polarization purity of the optical fields, our simulation shows that the first laser, second laser, telecom-photon cavity, and visible-photon cavity have polarization purities of 98\%, 99\%, 83\%, and 67\%, respectively. With these input parameters, our simulation shows that the Rb atom can emit telecom-visible photon pairs within $\sim60$ ns, as shown in Fig.~\ref{fig_Rb_emitter}(d). This is far shorter than the Rb state initialization time of roughly 1 $\mu$s. Hence we assume the Rb repetition rate to be 1 MHz.

\section{Thulium-doped Crystal Quantum Memory}
\label{sec_Tm}

\begin{figure}
\includegraphics[width=0.47\textwidth]{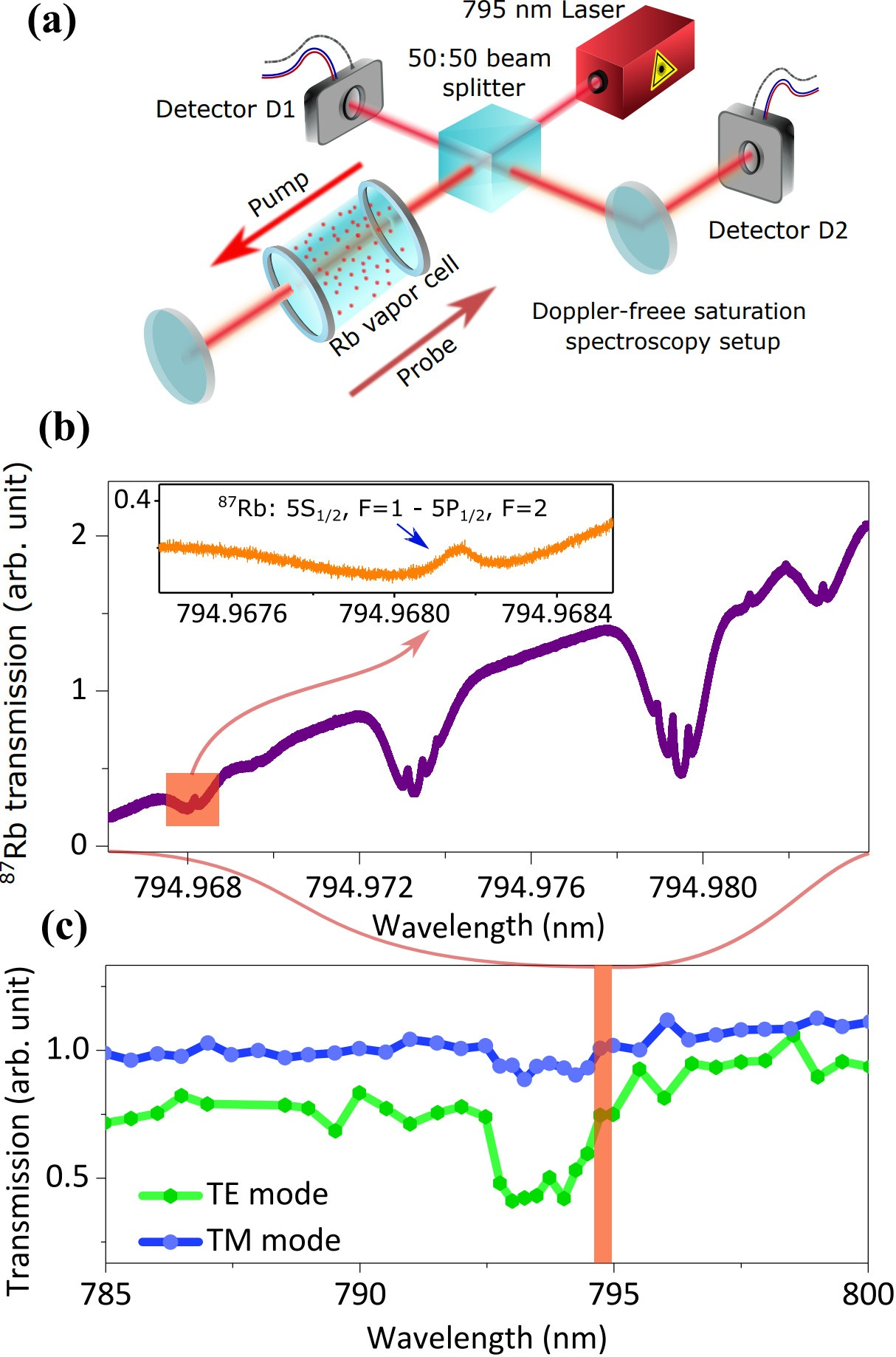}\centering
\caption{\label{fig:Rb_Tm_interface} (a) A simplified schematic of the Doppler-free saturation spectroscopy setup used to measure the absorption spectrum of rubidium. We use a laser with continuously sweeping frequency along with two light detectors (D1 to normalize the pump light, D2 to detect the probe light) to measure the absorption spectrum of the $^{87}$Rb vapor. (b) Absorption spectrum of $^{87}$Rb vapor. The inset shows an absorption peak corresponding to the [5$^2$S$_{1/2}$, $F=1$] $\leftrightarrow$ [5$^2$P$_{1/2}$, $F=2$] transition which is used in our single-spin photon transducer. (c) The absorption spectrum of Tm:LiNbO$_3$. The orange bar indicates the wavelength range in panel (b), identifying the spectral overlap between the Tm-memory absorption and Rb photon transducer emission.}
\end{figure}

In our proposal, multimode quantum memories (QMs) require an ensemble-based approach~\cite{QM1, jing2023ensemble}. Several possible methods exist, including electromagnetically induced transparency (EIT)~\cite{ma2017optical}, and photon-echo-related approaches~\cite{QM2_echo}, including the atomic frequency comb (AFC) protocol~\cite{AFC, jobez2016}. Relevant hardware include alkaline atoms~\cite{hot_Rb, Efficient_storage_1, Efficient_storage_2, Efficient_storage_3, Rb_Raman} and rare-earth-ion-doped crystals (REID crystals)~\cite{REI1, REI2, REI3, REI-Book}. While the transition of Rb vapor or laser-cooled Rb is naturally resonant with the emission wavelengths of photons from a single Rb atom, here we focus on the use of REID crystals. The reason is two-fold. Firstly, the multimode capacity of the AFC quantum memory protocol, which is widely used for quantum state storage in REID crystals~\cite{saglamyurek2016multiplexed, Businger2022}, does not depend on the optical depth of the material, unlike in the case of EIT~\cite{ma2017optical, nunn2008multimode}. Secondly, some Tm-doped crystals have matching resonance lines with $^{87}$Rb, as shown in Fig.~\ref{fig:Rb_Tm_interface} with the example of Tm-doped LiNbO$_3$(Tm:LiNbO$_3$): We performed Doppler-free saturation spectroscopy~\cite{preston1996doppler} of atomic $^{87}$Rb vapor using an experimental setup schematically shown in a simplified form in Fig.~\ref{fig:Rb_Tm_interface}(a). A pump beam propagates through the $^{87}$Rb vapor cell and turns into a probe beam after reflecting from a mirror passing through the $^{87}$Rb vapor cell again. The intensities of both beams are detected by two photon detectors, respectively. This gives the transmission of the $^{87}$Rb vapor as a function of the wavelength of the laser, i.e. the spectrum of the $^{87}$Rb atom as shown in Fig.~\ref{fig:Rb_Tm_interface}(b). On the other hand, the transmission spectrum of a cryogenically cooled ($\approx$ 600 mK) Tm:LiNbO$_3$ crystal is depicted in Fig.~\ref{fig:Rb_Tm_interface}(c), showing spectral overlap with the $^{87}$Rb spectrum. Specifically, we show an enlarged view of the [5$^2$S$_{1/2}$, $F=1$] $\leftrightarrow$ [5$^2$P$_{1/2}$, $F=2$] resonance line in the inset of Fig.~\ref{fig:Rb_Tm_interface}(b), which is used in our Rb single-spin photon transducer.

Tm:LiNbO$_3$ has been used in several implementations of QM for light~\cite{Multiplexing, Entanglement_and_nonlocality}. Ignoring spectral diffusion, which will be discussed below~\cite{bottger2006optical}, and assuming sufﬁcient optical depth, e.g. by using an impedance-matched cavity~\cite{Das:23}, the storage efficiency for photons at 795 nm wavelength is determined by the optical coherence time $T_2$ of its $^3H_6 \leftrightarrow$ $^3H_4$ transition. More precisely, the normalized retrieval rate decreases as $R_{\rm rtr}=\exp\left(-4t/T_2 \right)$, where $t$ is the storage time, yielding $t_{\rm 1/e}=T_2/4$. The coherence time itself is upper bounded by $T_2^{\rm max}=2T_1$ where $T_1$ is the lifetime of the excited state -- in the case of Tm:LiNbO$_3$ around 100 $\mu$s. Experimentally, a $T_2$ time of 117 $\mu$s  has been reported at a temperature of 810 mK~\cite{sinclair2017properties}, which limits $t_{\rm 1/e}$ to around 30$\mu$s. We note that other Tm-doped crystals with much longer coherence times are known. In particular, these include Tm:Y$_3$Ga$_5$O$_{12}$ (Tm:YGG), for which $T_2$=1.1 ms has been measured at 500 mK~\cite{Askarani2021}. We expect that the coherence time will approach the theoretical maximum of 2$T_1$=2.6 ms under optimized experimental conditions, in particular at lower temperatures and optimized magnetic fields.

In addition to the coherence time, spectral diffusion is another factor that can limit the possible storage time~\cite{sun2012optical, sinclair2021optical, saglamyurek2012conditional, sinclair2017properties, Askarani2021, bottger2006optical}. However, there are several ways to mitigate the effect of spectral diffusion, including so-called zero first-order Zeeman (ZEFOZ) transitions~\cite{mcauslan2012reducing, Davidson}, a reduction of the temperature~\cite{sinclair2017properties, sinclair2021optical}, or co-doping with other ions~\cite{Thiel_strain, Bottger_codoping}. For the simulations in this paper, we assume that the storage efficiency is only determined by the coherence time, which we furthermore assume to be given by its upper limit of 2$T_1$. In the case of Tm:YGG, we thus find $R_{\rm rtr}=\exp\left(-4t/2.6 {\rm ms}\right)$.

As shown in Fig. \ref{fig:Rb_Tm_interface}, the $^3H_6\leftrightarrow$ $^3H_4$ transition line of Tm:LiNbO$_3$ spectrally overlaps with the [5$^2$S$_{1/2}$, $F=1$] $\leftrightarrow$ [5$^2$P$_{1/2}$, $F=2$] line in $^{87}$Rb. However, the same line in Tm:YGG is slightly off-resonant (the Rb transition is centered at 794.97 nm and features a linewidth of 2$\pi\cdot$5.7 MHz while the line of a 1\% doped Tm:YGG crystal is centered at 795.32 nm with 56 GHz linewidth~\cite{thiel2014optical}). However, crystal engineering, e.g. by co-doping~\cite{Thiel_strain, Bottger_codoping}, may solve this problem. Experiments show that a linear increase of the inhomogeneous linewidth in Tm:Y$_3$Al$_5$O$_{12}$(Tm:YAG) can be induced, for example, by 24 GHz/Scandium$\%$~\cite{ferrier2018scandium} and by 3.6 GHz/Europium$\%$~\cite{zhang2020tailoring}. It is reasonable to assume that these methods work for Tm:YGG as well. While more work remains to be done, it therefore appears feasible to match the Rb line with the absorption line of a Tm-doped crystal, allowing the creation of quantum memory for light with the desired specifications.

\section{Simulated performance of the repeater chain}

\begin{figure}
\includegraphics[width=0.47\textwidth]{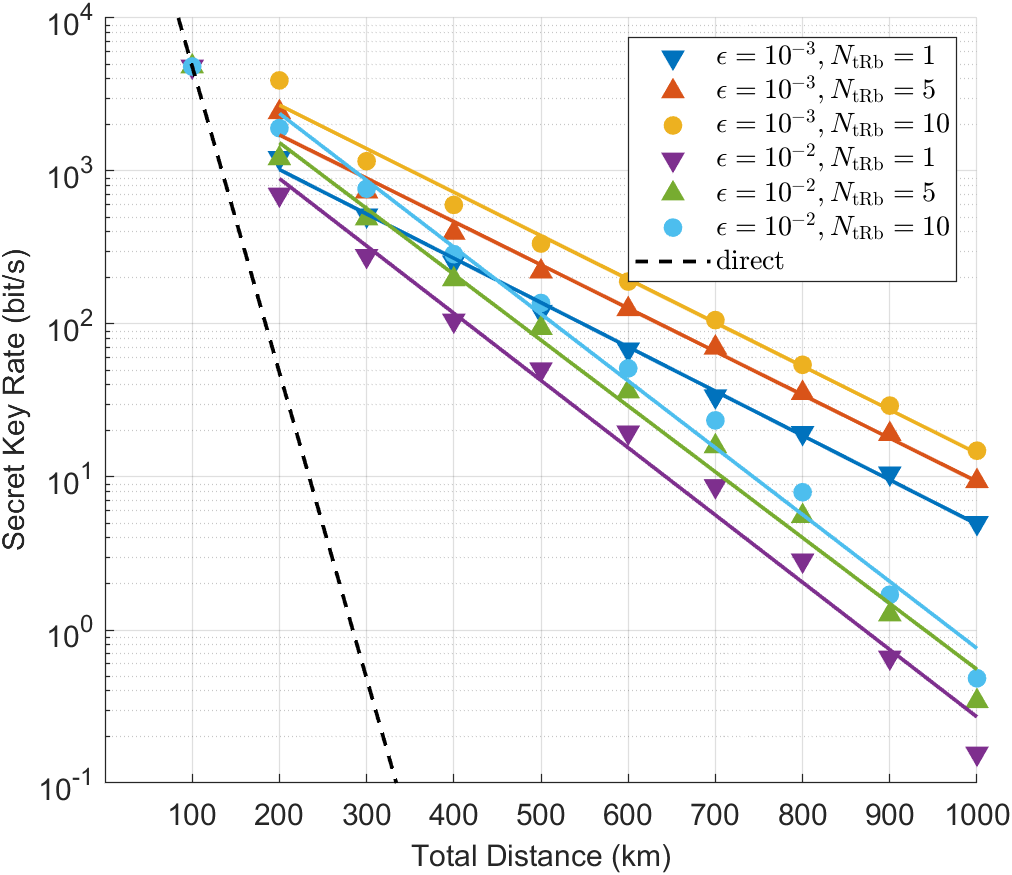}\centering
\caption{\label{fig_simulation} Average secrete key rate $R_{\rm SK}$ over the varied total distances $L$ for different $N_{\rm tRb}$, the number of Rb atoms for entanglement transfer at each side of the elementary segment, and $\epsilon$, the density matrix error induced by each entanglement swapping. The dashed black line presents the secret key rate with no repeaters but only a direct link between the two end notes. The solid lines are linear fittings for the data with the corresponding color. The data points are calculated based on the optimal number of repeaters for each case as shown in Appendix~E and the cutoff time is unified to be $t_{\rm cut}=10$ ms. The solid lines are the linear fittings for the data points where the ones at $L=100$ km are excluded because the direct link is optimal for that case. The code of this repeater chain simulation is available in ref~\cite{codeCollection}.}
\end{figure}

To assess the performance of the quantum repeater chain, we conduct an in-depth simulation to determine the secret key rate achievable when applied to quantum key distribution (QKD). Through a Monte-Carlo simulation, we track the successful entanglement generation, transfer, and swapping events throughout the repeater chain, allowing for accurate modeling of finite memory time effects on the quality of the entanglement. For each data point, we did 100 independent simulations. In each simulation, we simulate the time, $T_i$, it takes before successful end-to-end entanglement has been established 1000 times. From this, we estimate the average rate of entanglement distribution as $r_i=1000/T_i$ for the $i$'th simulation. This is then further averaged over the 100 independent simulations such that our final estimate of the average entanglement distribution rate is $R_{\text{suc}}=\sum_{i=1}^{100}r_i/100$.

We trace the evolution of the density matrices of each entangled pair including the decoherence due to imperfect storage in the Rb atom at the repeater nodes, which we model as a depolarizing channel. As a result, we obtain 100,000 density matrices of the end-to-end entangled pair from the 100 independent simulations, each with 1000 successful events. From these, we derive the average qubit error rates (QBERs) of Z and X type errors, $Q_{{\rm Z}}$ and $Q_{{\rm X}}$, from which, based on the BB84 protocol~\cite{Bennett2014}, we estimate the raw secrete key rate as
\begin{equation}
 R_{{\rm SK}}=\max\left(0,\ 1-H\left(Q_{{\rm X}}\right)-H\left(Q_{{\rm Z}}\right)\right)\times R_{{\rm suc}},
\label{Rsk}   
\end{equation}
where $H(\cdot)$ is the binary entropy function. Please note that, in the original BB84 protocol, there is an overall factor of 1/2 from the random basis choice. This can, however, be circumvented without loss of security in the asymptotic limit of long keys~\cite{Lo2005, Scarani2009} that we consider here such that this factor is omitted.

As mentioned in section~\ref{section_Rb}, we assume that the Rb emitter has a repetition rate of $\nu_{\rm Rb}=$ 1 MHz, producing entanglement pairs of fidelity 0.98. The fiber attenuation rate for the telecom photon is 0.2 dB/km and the speed of light in the fiber is $\rm{c}=2\times 10^5$ km/s. In addition, the single photon detection efficiency is set to be 0.99.  As mentioned in section~\ref{sec_Tm}, the retrieving rate of the Tm-memory as a function of storage time $t$ is $R_{\rm rtr}=\exp\left(-4t/2.6 {\rm ms}\right)$. The entanglement transfer is chosen to have an overall success rate of 0.95 including optical circuit loss and imperfect photon detection efficiency. The coherence time of the hyperfine levels of the individually trapped Rb atoms is set to 1s. The local entanglement swap has an operation time of 200 ns and a success rate of 0.92 which is a combination of 0.97 efficiency of emitting a photon from one of the Rb-atoms (assuming a cooperativity of 30) and a total efficiency of 0.95 for entangling with the other atom following the entanglement transfer process. To examine the repeater performance under different entanglement swapping error rates, we simulated two cases with the error probability being $\epsilon=10^{-2}$ and $\epsilon=10^{-3}$. Thus, with probability $1-\epsilon$, the entanglement swap operation is perfect while with probability $\epsilon$, the swapped state is completely mixed.

We assume that the two end nodes can perfectly store the successfully entangled pairs or directly measure the pairs without needing storage depending on the QKD protocols. Thus we do not consider imperfect storage or limited memory capabilities at the end nodes but include this in all repeater nodes. Because of this, we introduced a cutoff time so that an entangled pair of Rb atoms is discarded after an idling time of $t_{\rm cut}$.

In addition, instead of the case with only one Rb photon transducer for entanglement transfer at each end of each segment as shown in Fig.~\ref{fig_structure}, we also simulate cases where multiple ($N_{\rm tRb}$) Rb photon transducers are used. This provides a second level of multiplexing for the entanglement swapping.

Fig.~\ref{fig_simulation} shows the simulated average secret key rates $R_{{\rm SK}}$ as a function of the total distance $L$ for various $N_{\rm tRb}$ and errors $\epsilon$ due to imperfect entanglement swapping. We optimize the number of repeaters in the chain $N_{\rm rep}$ and the cutoff time $t_{\rm cut}$ to achieve the highest secret key rate per elementary segment, i.e., $R_{\rm SK}/N_{\rm seg}$ with $N_{\rm seg}=N_{\rm rep}+1$. We find that for an Rb spin coherence time of 1 s, a cutoff time of $t_{\rm cut} = 10$ ms is approximately optimum in all simulations. The optimum number of repeater stations $N_{\rm rep}$ used for each data point in Fig.~\ref{fig_simulation} can be found in Appendix~E.

Our simulation shows that secret key rates on the order of 10 bit/s and 1 bit/s can be achieved over 1000 km for entanglement swapping error rates $\epsilon=10^{-3}$ and $\epsilon=10^{-2}$, respectively. We have also assumed that the Tm-memories have sufficient multiplexing capacity to ensure the continued operation of the repeater. This means that the number of memory modes needs to be $N_{\rm mode}=\nu_{\rm Rb}\frac{L}{N_{\rm seg}\rm{c}}$. Based on the total distances and the optimum number of repeaters, $N_{\rm mode}$ ranges from 167 to 625 in the cases we simulated. The entanglement distribution rate could be further increased by increasing the number of Rb-atoms both in the repeater nodes and if more Rb-atoms were used in the entanglement generation step to increase the repetition rate of the entangled photon pairs.

\section{Discussion}
In summary, we have proposed a hybrid quantum repeater employing single-spin photon transducers and ensemble-based photonic memories to achieve high-rate entanglement distribution over large distances. The hybrid architecture directs the massive multiplexing necessary for battling transmission loss in the optical fibers to the ensemble-based memories, while efficient and near-ideal photon pair generation and entanglement swapping are enabled by the single-spin photon transducers. Furthermore, we provided a specific example utilizing Rb atoms coupled to nanophotonic cavities as single-spin photon transducers and Tm-doped crystal as ensemble-based memories for massive multiplexing. 

The overall framework presented here is relevant to other combinations of hardware besides the specific example analyzed in this work. Efficient single-spin photon transducers can be realized with diamond defect centers~\cite{knaut2023} and quantum dot systems~\cite{Huber2018}, which can be matched with other ensemble-based memories based on AFC~\cite{AFC},  Raman~\cite{PhysRevApplied.18.024036} or EIT~\cite{ma2017optical} storage using impurity-doped crystals~\cite{guo2023rare, REI-Book} or atoms, either laser-cooled or at room temperature. We note that different hardware combinations may require frequency conversion to be compatible, which can be achieved through standard techniques based on non-linear waveguides~\cite{Tanzilli2005,Tchebotareva2019,Iuliano2023,knaut2023}. 

While combining different hardware is arguably more complex than a single hardware repeater, the overall requirements for reaching high-rate entanglement distribution may be substantially relaxed as demonstrated in this work. We thus believe that further investigations of hybrid repeater architectures where ensemble-based memories are combined with single quantum emitters are a promising direction for future work. In particular, employing more complex operations such as entanglement purification techniques~\cite{Kalb2017} to boost the fidelity could be envisioned through the quantum logic enabled by the single-spin photon transducers. 

\begin{acknowledgments}
J.B. and F.G. acknowledge funding from the NWO Gravitation Program Quantum Software Consortium (Project QSC No. 024.003.037). J.B. acknowledges support from The AWS Quantum Discovery Fund at the Harvard Quantum Initiative. H.B. and S.G.M. gratefully acknowledge funding from the NSF
QLCI for Hybrid Quantum Architectures and Networks
(NSF award 2016136), and the NSF Quantum Interconnects
Challenge for Transformational Advances in Quantum
Systems (NSF award 2138068). W.T. acknowledges funding through the Netherlands Organization for Scientific Research, and the European Union's Horizon 2020 Research and Innovation Program under Grant Agreement No. 820445 and Project Name Quantum Internet Alliance.
\end{acknowledgments}

\bibliography{references}

\begin{thebibliography}{84}%
\makeatletter
\providecommand \@ifxundefined [1]{%
 \@ifx{#1\undefined}
}%
\providecommand \@ifnum [1]{%
 \ifnum #1\expandafter \@firstoftwo
 \else \expandafter \@secondoftwo
 \fi
}%
\providecommand \@ifx [1]{%
 \ifx #1\expandafter \@firstoftwo
 \else \expandafter \@secondoftwo
 \fi
}%
\providecommand \natexlab [1]{#1}%
\providecommand \enquote  [1]{``#1''}%
\providecommand \bibnamefont  [1]{#1}%
\providecommand \bibfnamefont [1]{#1}%
\providecommand \citenamefont [1]{#1}%
\providecommand \href@noop [0]{\@secondoftwo}%
\providecommand \href [0]{\begingroup \@sanitize@url \@href}%
\providecommand \@href[1]{\@@startlink{#1}\@@href}%
\providecommand \@@href[1]{\endgroup#1\@@endlink}%
\providecommand \@sanitize@url [0]{\catcode `\\12\catcode `\$12\catcode
  `\&12\catcode `\#12\catcode `\^12\catcode `\_12\catcode `\%12\relax}%
\providecommand \@@startlink[1]{}%
\providecommand \@@endlink[0]{}%
\providecommand \url  [0]{\begingroup\@sanitize@url \@url }%
\providecommand \@url [1]{\endgroup\@href {#1}{\urlprefix }}%
\providecommand \urlprefix  [0]{URL }%
\providecommand \Eprint [0]{\href }%
\providecommand \doibase [0]{https://doi.org/}%
\providecommand \selectlanguage [0]{\@gobble}%
\providecommand \bibinfo  [0]{\@secondoftwo}%
\providecommand \bibfield  [0]{\@secondoftwo}%
\providecommand \translation [1]{[#1]}%
\providecommand \BibitemOpen [0]{}%
\providecommand \bibitemStop [0]{}%
\providecommand \bibitemNoStop [0]{.\EOS\space}%
\providecommand \EOS [0]{\spacefactor3000\relax}%
\providecommand \BibitemShut  [1]{\csname bibitem#1\endcsname}%
\let\auto@bib@innerbib\@empty
\bibitem [{\citenamefont {Kimble}(2008)}]{Kimble2008}%
  \BibitemOpen
  \bibfield  {author} {\bibinfo {author} {\bibfnamefont {H.~J.}\ \bibnamefont
  {Kimble}},\ }\bibfield  {title} {\bibinfo {title} {The quantum internet},\
  }\href {https://doi.org/10.1038/nature07127} {\bibfield  {journal} {\bibinfo
  {journal} {Nature}\ }\textbf {\bibinfo {volume} {453}},\ \bibinfo {pages}
  {1023} (\bibinfo {year} {2008})}\BibitemShut {NoStop}%
\bibitem [{\citenamefont {Wehner}\ \emph {et~al.}(2018)\citenamefont {Wehner},
  \citenamefont {Elkouss},\ and\ \citenamefont {Hanson}}]{Wehner2018}%
  \BibitemOpen
  \bibfield  {author} {\bibinfo {author} {\bibfnamefont {S.}~\bibnamefont
  {Wehner}}, \bibinfo {author} {\bibfnamefont {D.}~\bibnamefont {Elkouss}},\
  and\ \bibinfo {author} {\bibfnamefont {R.}~\bibnamefont {Hanson}},\
  }\bibfield  {title} {\bibinfo {title} {Quantum internet: A vision for the
  road ahead},\ }\href {https://doi.org/10.1126/science.aam9288} {\bibfield
  {journal} {\bibinfo  {journal} {Science}\ }\textbf {\bibinfo {volume}
  {362}},\ \bibinfo {pages} {eaam9288} (\bibinfo {year} {2018})}\BibitemShut
  {NoStop}%
\bibitem [{\citenamefont {Munro}\ \emph {et~al.}(2015)\citenamefont {Munro},
  \citenamefont {Azuma}, \citenamefont {Tamaki},\ and\ \citenamefont
  {Nemoto}}]{Munro2015}%
  \BibitemOpen
  \bibfield  {author} {\bibinfo {author} {\bibfnamefont {W.~J.}\ \bibnamefont
  {Munro}}, \bibinfo {author} {\bibfnamefont {K.}~\bibnamefont {Azuma}},
  \bibinfo {author} {\bibfnamefont {K.}~\bibnamefont {Tamaki}},\ and\ \bibinfo
  {author} {\bibfnamefont {K.}~\bibnamefont {Nemoto}},\ }\bibfield  {title}
  {\bibinfo {title} {Inside quantum repeaters},\ }\href
  {https://doi.org/10.1109/JSTQE.2015.2392076} {\bibfield  {journal} {\bibinfo
  {journal} {IEEE Journal of Selected Topics in Quantum Electronics}\ }\textbf
  {\bibinfo {volume} {21}},\ \bibinfo {pages} {78} (\bibinfo {year}
  {2015})}\BibitemShut {NoStop}%
\bibitem [{\citenamefont {Azuma}\ \emph {et~al.}(2023)\citenamefont {Azuma},
  \citenamefont {Economou}, \citenamefont {Elkouss}, \citenamefont {Hilaire},
  \citenamefont {Jiang}, \citenamefont {Lo},\ and\ \citenamefont
  {Tzitrin}}]{azuma2023_RMP}%
  \BibitemOpen
  \bibfield  {author} {\bibinfo {author} {\bibfnamefont {K.}~\bibnamefont
  {Azuma}}, \bibinfo {author} {\bibfnamefont {S.~E.}\ \bibnamefont {Economou}},
  \bibinfo {author} {\bibfnamefont {D.}~\bibnamefont {Elkouss}}, \bibinfo
  {author} {\bibfnamefont {P.}~\bibnamefont {Hilaire}}, \bibinfo {author}
  {\bibfnamefont {L.}~\bibnamefont {Jiang}}, \bibinfo {author} {\bibfnamefont
  {H.-K.}\ \bibnamefont {Lo}},\ and\ \bibinfo {author} {\bibfnamefont
  {I.}~\bibnamefont {Tzitrin}},\ }\bibfield  {title} {\bibinfo {title} {Quantum
  repeaters: From quantum networks to the quantum internet},\ }\href
  {https://doi.org/10.1103/RevModPhys.95.045006} {\bibfield  {journal}
  {\bibinfo  {journal} {Rev. Mod. Phys.}\ }\textbf {\bibinfo {volume} {95}},\
  \bibinfo {pages} {045006} (\bibinfo {year} {2023})}\BibitemShut {NoStop}%
\bibitem [{\citenamefont {Hermans}\ \emph {et~al.}(2022)\citenamefont
  {Hermans}, \citenamefont {Pompili}, \citenamefont {Beukers}, \citenamefont
  {Baier}, \citenamefont {Borregaard},\ and\ \citenamefont
  {Hanson}}]{Hermans2022}%
  \BibitemOpen
  \bibfield  {author} {\bibinfo {author} {\bibfnamefont {S.~L.~N.}\
  \bibnamefont {Hermans}}, \bibinfo {author} {\bibfnamefont {M.}~\bibnamefont
  {Pompili}}, \bibinfo {author} {\bibfnamefont {H.~K.~C.}\ \bibnamefont
  {Beukers}}, \bibinfo {author} {\bibfnamefont {S.}~\bibnamefont {Baier}},
  \bibinfo {author} {\bibfnamefont {J.}~\bibnamefont {Borregaard}},\ and\
  \bibinfo {author} {\bibfnamefont {R.}~\bibnamefont {Hanson}},\ }\bibfield
  {title} {\bibinfo {title} {Qubit teleportation between non-neighbouring nodes
  in a quantum network},\ }\href {https://doi.org/10.1038/s41586-022-04697-y}
  {\bibfield  {journal} {\bibinfo  {journal} {Nature}\ }\textbf {\bibinfo
  {volume} {605}},\ \bibinfo {pages} {663} (\bibinfo {year}
  {2022})}\BibitemShut {NoStop}%
\bibitem [{\citenamefont {Knaut}\ \emph {et~al.}(2023)\citenamefont {Knaut},
  \citenamefont {Suleymanzade}, \citenamefont {Wei}, \citenamefont {Assumpcao},
  \citenamefont {Stas}, \citenamefont {Huan}, \citenamefont {Machielse},
  \citenamefont {Knall}, \citenamefont {Sutula}, \citenamefont {Baranes},
  \citenamefont {Sinclair}, \citenamefont {De-Eknamkul}, \citenamefont
  {Levonian}, \citenamefont {Bhaskar}, \citenamefont {Park}, \citenamefont
  {Lončar},\ and\ \citenamefont {Lukin}}]{knaut2023}%
  \BibitemOpen
  \bibfield  {author} {\bibinfo {author} {\bibfnamefont {C.~M.}\ \bibnamefont
  {Knaut}}, \bibinfo {author} {\bibfnamefont {A.}~\bibnamefont {Suleymanzade}},
  \bibinfo {author} {\bibfnamefont {Y.-C.}\ \bibnamefont {Wei}}, \bibinfo
  {author} {\bibfnamefont {D.~R.}\ \bibnamefont {Assumpcao}}, \bibinfo {author}
  {\bibfnamefont {P.-J.}\ \bibnamefont {Stas}}, \bibinfo {author}
  {\bibfnamefont {Y.~Q.}\ \bibnamefont {Huan}}, \bibinfo {author}
  {\bibfnamefont {B.}~\bibnamefont {Machielse}}, \bibinfo {author}
  {\bibfnamefont {E.~N.}\ \bibnamefont {Knall}}, \bibinfo {author}
  {\bibfnamefont {M.}~\bibnamefont {Sutula}}, \bibinfo {author} {\bibfnamefont
  {G.}~\bibnamefont {Baranes}}, \bibinfo {author} {\bibfnamefont
  {N.}~\bibnamefont {Sinclair}}, \bibinfo {author} {\bibfnamefont
  {C.}~\bibnamefont {De-Eknamkul}}, \bibinfo {author} {\bibfnamefont {D.~S.}\
  \bibnamefont {Levonian}}, \bibinfo {author} {\bibfnamefont {M.~K.}\
  \bibnamefont {Bhaskar}}, \bibinfo {author} {\bibfnamefont {H.}~\bibnamefont
  {Park}}, \bibinfo {author} {\bibfnamefont {M.}~\bibnamefont {Lončar}},\ and\
  \bibinfo {author} {\bibfnamefont {M.~D.}\ \bibnamefont {Lukin}},\ }\href@noop
  {} {\bibinfo {title} {Entanglement of nanophotonic quantum memory nodes in a
  telecommunication network}} (\bibinfo {year} {2023}),\ \Eprint
  {https://arxiv.org/abs/2310.01316} {arXiv:2310.01316 [quant-ph]} \BibitemShut
  {NoStop}%
\bibitem [{\citenamefont {Heller}\ \emph {et~al.}(2020)\citenamefont {Heller},
  \citenamefont {Farrera}, \citenamefont {Heinze},\ and\ \citenamefont
  {de~Riedmatten}}]{Heller2020}%
  \BibitemOpen
  \bibfield  {author} {\bibinfo {author} {\bibfnamefont {L.}~\bibnamefont
  {Heller}}, \bibinfo {author} {\bibfnamefont {P.}~\bibnamefont {Farrera}},
  \bibinfo {author} {\bibfnamefont {G.}~\bibnamefont {Heinze}},\ and\ \bibinfo
  {author} {\bibfnamefont {H.}~\bibnamefont {de~Riedmatten}},\ }\bibfield
  {title} {\bibinfo {title} {Cold-atom temporally multiplexed quantum memory
  with cavity-enhanced noise suppression},\ }\href
  {https://doi.org/10.1103/PhysRevLett.124.210504} {\bibfield  {journal}
  {\bibinfo  {journal} {Phys. Rev. Lett.}\ }\textbf {\bibinfo {volume} {124}},\
  \bibinfo {pages} {210504} (\bibinfo {year} {2020})}\BibitemShut {NoStop}%
\bibitem [{\citenamefont {Pu}\ \emph {et~al.}(2021)\citenamefont {Pu},
  \citenamefont {Zhang}, \citenamefont {Wu}, \citenamefont {Jiang},
  \citenamefont {Chang}, \citenamefont {Li},\ and\ \citenamefont
  {Duan}}]{Pu2021}%
  \BibitemOpen
  \bibfield  {author} {\bibinfo {author} {\bibfnamefont {Y.-F.}\ \bibnamefont
  {Pu}}, \bibinfo {author} {\bibfnamefont {S.}~\bibnamefont {Zhang}}, \bibinfo
  {author} {\bibfnamefont {Y.-K.}\ \bibnamefont {Wu}}, \bibinfo {author}
  {\bibfnamefont {N.}~\bibnamefont {Jiang}}, \bibinfo {author} {\bibfnamefont
  {W.}~\bibnamefont {Chang}}, \bibinfo {author} {\bibfnamefont
  {C.}~\bibnamefont {Li}},\ and\ \bibinfo {author} {\bibfnamefont {L.-M.}\
  \bibnamefont {Duan}},\ }\bibfield  {title} {\bibinfo {title} {Experimental
  demonstration of memory-enhanced scaling for entanglement connection of
  quantum repeater segments},\ }\href
  {https://doi.org/10.1038/s41566-021-00764-4} {\bibfield  {journal} {\bibinfo
  {journal} {Nature Photonics}\ }\textbf {\bibinfo {volume} {15}},\ \bibinfo
  {pages} {374} (\bibinfo {year} {2021})}\BibitemShut {NoStop}%
\bibitem [{\citenamefont {Askarani}\ \emph {et~al.}(2021)\citenamefont
  {Askarani}, \citenamefont {Das}, \citenamefont {Davidson}, \citenamefont
  {Amaral}, \citenamefont {Sinclair}, \citenamefont {Slater}, \citenamefont
  {Marzban}, \citenamefont {Thiel}, \citenamefont {Cone}, \citenamefont
  {Oblak},\ and\ \citenamefont {Tittel}}]{Askarani2021}%
  \BibitemOpen
  \bibfield  {author} {\bibinfo {author} {\bibfnamefont {M.~F.}\ \bibnamefont
  {Askarani}}, \bibinfo {author} {\bibfnamefont {A.}~\bibnamefont {Das}},
  \bibinfo {author} {\bibfnamefont {J.~H.}\ \bibnamefont {Davidson}}, \bibinfo
  {author} {\bibfnamefont {G.~C.}\ \bibnamefont {Amaral}}, \bibinfo {author}
  {\bibfnamefont {N.}~\bibnamefont {Sinclair}}, \bibinfo {author}
  {\bibfnamefont {J.~A.}\ \bibnamefont {Slater}}, \bibinfo {author}
  {\bibfnamefont {S.}~\bibnamefont {Marzban}}, \bibinfo {author} {\bibfnamefont
  {C.~W.}\ \bibnamefont {Thiel}}, \bibinfo {author} {\bibfnamefont {R.~L.}\
  \bibnamefont {Cone}}, \bibinfo {author} {\bibfnamefont {D.}~\bibnamefont
  {Oblak}},\ and\ \bibinfo {author} {\bibfnamefont {W.}~\bibnamefont
  {Tittel}},\ }\bibfield  {title} {\bibinfo {title} {Long-lived solid-state
  optical memory for high-rate quantum repeaters},\ }\href
  {https://doi.org/10.1103/PhysRevLett.127.220502} {\bibfield  {journal}
  {\bibinfo  {journal} {Phys. Rev. Lett.}\ }\textbf {\bibinfo {volume} {127}},\
  \bibinfo {pages} {220502} (\bibinfo {year} {2021})}\BibitemShut {NoStop}%
\bibitem [{\citenamefont {Bock}\ \emph {et~al.}(2018)\citenamefont {Bock},
  \citenamefont {Eich}, \citenamefont {Kucera}, \citenamefont {Kreis},
  \citenamefont {Lenhard}, \citenamefont {Becher},\ and\ \citenamefont
  {Eschner}}]{Bock2018}%
  \BibitemOpen
  \bibfield  {author} {\bibinfo {author} {\bibfnamefont {M.}~\bibnamefont
  {Bock}}, \bibinfo {author} {\bibfnamefont {P.}~\bibnamefont {Eich}}, \bibinfo
  {author} {\bibfnamefont {S.}~\bibnamefont {Kucera}}, \bibinfo {author}
  {\bibfnamefont {M.}~\bibnamefont {Kreis}}, \bibinfo {author} {\bibfnamefont
  {A.}~\bibnamefont {Lenhard}}, \bibinfo {author} {\bibfnamefont
  {C.}~\bibnamefont {Becher}},\ and\ \bibinfo {author} {\bibfnamefont
  {J.}~\bibnamefont {Eschner}},\ }\bibfield  {title} {\bibinfo {title}
  {High-fidelity entanglement between a trapped ion and a telecom photon via
  quantum frequency conversion},\ }\href
  {https://doi.org/10.1038/s41467-018-04341-2} {\bibfield  {journal} {\bibinfo
  {journal} {Nature Communications}\ }\textbf {\bibinfo {volume} {9}},\
  \bibinfo {pages} {1998} (\bibinfo {year} {2018})}\BibitemShut {NoStop}%
\bibitem [{\citenamefont {Krutyanskiy}\ \emph {et~al.}(2023)\citenamefont
  {Krutyanskiy}, \citenamefont {Canteri}, \citenamefont {Meraner},
  \citenamefont {Bate}, \citenamefont {Krcmarsky}, \citenamefont {Schupp},
  \citenamefont {Sangouard},\ and\ \citenamefont {Lanyon}}]{Krutyanskiy2023}%
  \BibitemOpen
  \bibfield  {author} {\bibinfo {author} {\bibfnamefont {V.}~\bibnamefont
  {Krutyanskiy}}, \bibinfo {author} {\bibfnamefont {M.}~\bibnamefont
  {Canteri}}, \bibinfo {author} {\bibfnamefont {M.}~\bibnamefont {Meraner}},
  \bibinfo {author} {\bibfnamefont {J.}~\bibnamefont {Bate}}, \bibinfo {author}
  {\bibfnamefont {V.}~\bibnamefont {Krcmarsky}}, \bibinfo {author}
  {\bibfnamefont {J.}~\bibnamefont {Schupp}}, \bibinfo {author} {\bibfnamefont
  {N.}~\bibnamefont {Sangouard}},\ and\ \bibinfo {author} {\bibfnamefont
  {B.~P.}\ \bibnamefont {Lanyon}},\ }\bibfield  {title} {\bibinfo {title}
  {Telecom-wavelength quantum repeater node based on a trapped-ion processor},\
  }\href {https://doi.org/10.1103/PhysRevLett.130.213601} {\bibfield  {journal}
  {\bibinfo  {journal} {Phys. Rev. Lett.}\ }\textbf {\bibinfo {volume} {130}},\
  \bibinfo {pages} {213601} (\bibinfo {year} {2023})}\BibitemShut {NoStop}%
\bibitem [{\citenamefont {Neuwirth}\ \emph {et~al.}(2021)\citenamefont
  {Neuwirth}, \citenamefont {Basset}, \citenamefont {Rota}, \citenamefont
  {Roccia}, \citenamefont {Schimpf}, \citenamefont {J{\"o}ns}, \citenamefont
  {Rastelli},\ and\ \citenamefont {Trotta}}]{Neuwirth2021}%
  \BibitemOpen
  \bibfield  {author} {\bibinfo {author} {\bibfnamefont {J.}~\bibnamefont
  {Neuwirth}}, \bibinfo {author} {\bibfnamefont {F.~B.}\ \bibnamefont
  {Basset}}, \bibinfo {author} {\bibfnamefont {M.~B.}\ \bibnamefont {Rota}},
  \bibinfo {author} {\bibfnamefont {E.}~\bibnamefont {Roccia}}, \bibinfo
  {author} {\bibfnamefont {C.}~\bibnamefont {Schimpf}}, \bibinfo {author}
  {\bibfnamefont {K.~D.}\ \bibnamefont {J{\"o}ns}}, \bibinfo {author}
  {\bibfnamefont {A.}~\bibnamefont {Rastelli}},\ and\ \bibinfo {author}
  {\bibfnamefont {R.}~\bibnamefont {Trotta}},\ }\bibfield  {title} {\bibinfo
  {title} {Quantum dot technology for quantum repeaters: from entangled photon
  generation toward the integration with quantum memories},\ }\href
  {https://doi.org/10.1088/2633-4356/ac3d14} {\bibfield  {journal} {\bibinfo
  {journal} {Materials for Quantum Technology}\ }\textbf {\bibinfo {volume}
  {1}},\ \bibinfo {pages} {043001} (\bibinfo {year} {2021})}\BibitemShut
  {NoStop}%
\bibitem [{\citenamefont {Sangouard}\ \emph {et~al.}(2011)\citenamefont
  {Sangouard}, \citenamefont {Simon}, \citenamefont {de~Riedmatten},\ and\
  \citenamefont {Gisin}}]{Sangouard2009}%
  \BibitemOpen
  \bibfield  {author} {\bibinfo {author} {\bibfnamefont {N.}~\bibnamefont
  {Sangouard}}, \bibinfo {author} {\bibfnamefont {C.}~\bibnamefont {Simon}},
  \bibinfo {author} {\bibfnamefont {H.}~\bibnamefont {de~Riedmatten}},\ and\
  \bibinfo {author} {\bibfnamefont {N.}~\bibnamefont {Gisin}},\ }\bibfield
  {title} {\bibinfo {title} {Quantum repeaters based on atomic ensembles and
  linear optics},\ }\href {https://doi.org/10.1103/RevModPhys.83.33} {\bibfield
   {journal} {\bibinfo  {journal} {Rev. Mod. Phys.}\ }\textbf {\bibinfo
  {volume} {83}},\ \bibinfo {pages} {33} (\bibinfo {year} {2011})}\BibitemShut
  {NoStop}%
\bibitem [{\citenamefont {Nemoto}\ \emph {et~al.}(2016)\citenamefont {Nemoto},
  \citenamefont {Trupke}, \citenamefont {Devitt}, \citenamefont
  {Scharfenberger}, \citenamefont {Buczak}, \citenamefont {Schmiedmayer},\ and\
  \citenamefont {Munro}}]{Nemoto2016}%
  \BibitemOpen
  \bibfield  {author} {\bibinfo {author} {\bibfnamefont {K.}~\bibnamefont
  {Nemoto}}, \bibinfo {author} {\bibfnamefont {M.}~\bibnamefont {Trupke}},
  \bibinfo {author} {\bibfnamefont {S.~J.}\ \bibnamefont {Devitt}}, \bibinfo
  {author} {\bibfnamefont {B.}~\bibnamefont {Scharfenberger}}, \bibinfo
  {author} {\bibfnamefont {K.}~\bibnamefont {Buczak}}, \bibinfo {author}
  {\bibfnamefont {J.}~\bibnamefont {Schmiedmayer}},\ and\ \bibinfo {author}
  {\bibfnamefont {W.~J.}\ \bibnamefont {Munro}},\ }\bibfield  {title} {\bibinfo
  {title} {Photonic quantum networks formed from nv- centers},\ }\href@noop {}
  {\bibfield  {journal} {\bibinfo  {journal} {Scientific reports}\ }\textbf
  {\bibinfo {volume} {6}},\ \bibinfo {pages} {26284} (\bibinfo {year}
  {2016})}\BibitemShut {NoStop}%
\bibitem [{\citenamefont {Borregaard}\ \emph {et~al.}(2020)\citenamefont
  {Borregaard}, \citenamefont {Pichler}, \citenamefont {Schr\"oder},
  \citenamefont {Lukin}, \citenamefont {Lodahl},\ and\ \citenamefont
  {S\o{}rensen}}]{Borregaard2020}%
  \BibitemOpen
  \bibfield  {author} {\bibinfo {author} {\bibfnamefont {J.}~\bibnamefont
  {Borregaard}}, \bibinfo {author} {\bibfnamefont {H.}~\bibnamefont {Pichler}},
  \bibinfo {author} {\bibfnamefont {T.}~\bibnamefont {Schr\"oder}}, \bibinfo
  {author} {\bibfnamefont {M.~D.}\ \bibnamefont {Lukin}}, \bibinfo {author}
  {\bibfnamefont {P.}~\bibnamefont {Lodahl}},\ and\ \bibinfo {author}
  {\bibfnamefont {A.~S.}\ \bibnamefont {S\o{}rensen}},\ }\bibfield  {title}
  {\bibinfo {title} {One-way quantum repeater based on near-deterministic
  photon-emitter interfaces},\ }\href
  {https://doi.org/10.1103/PhysRevX.10.021071} {\bibfield  {journal} {\bibinfo
  {journal} {Phys. Rev. X}\ }\textbf {\bibinfo {volume} {10}},\ \bibinfo
  {pages} {021071} (\bibinfo {year} {2020})}\BibitemShut {NoStop}%
\bibitem [{\citenamefont {Sharman}\ \emph {et~al.}(2021)\citenamefont
  {Sharman}, \citenamefont {Kimiaee~Asadi}, \citenamefont {Wein},\ and\
  \citenamefont {Simon}}]{Sharman2021}%
  \BibitemOpen
  \bibfield  {author} {\bibinfo {author} {\bibfnamefont {K.}~\bibnamefont
  {Sharman}}, \bibinfo {author} {\bibfnamefont {F.}~\bibnamefont
  {Kimiaee~Asadi}}, \bibinfo {author} {\bibfnamefont {S.~C.}\ \bibnamefont
  {Wein}},\ and\ \bibinfo {author} {\bibfnamefont {C.}~\bibnamefont {Simon}},\
  }\bibfield  {title} {\bibinfo {title} {Quantum repeaters based on individual
  electron spins and nuclear-spin-ensemble memories in quantum dots},\ }\href
  {https://doi.org/10.22331/q-2021-11-02-570} {\bibfield  {journal} {\bibinfo
  {journal} {{Quantum}}\ }\textbf {\bibinfo {volume} {5}},\ \bibinfo {pages}
  {570} (\bibinfo {year} {2021})}\BibitemShut {NoStop}%
\bibitem [{\citenamefont {Duan}\ \emph {et~al.}(2001)\citenamefont {Duan},
  \citenamefont {Lukin}, \citenamefont {Cirac},\ and\ \citenamefont
  {Zoller}}]{Duan2001}%
  \BibitemOpen
  \bibfield  {author} {\bibinfo {author} {\bibfnamefont {L.~M.}\ \bibnamefont
  {Duan}}, \bibinfo {author} {\bibfnamefont {M.~D.}\ \bibnamefont {Lukin}},
  \bibinfo {author} {\bibfnamefont {J.~I.}\ \bibnamefont {Cirac}},\ and\
  \bibinfo {author} {\bibfnamefont {P.}~\bibnamefont {Zoller}},\ }\bibfield
  {title} {\bibinfo {title} {Long-distance quantum communication with atomic
  ensembles and linear optics},\ }\href {https://doi.org/10.1038/35106500}
  {\bibfield  {journal} {\bibinfo  {journal} {Nature}\ }\textbf {\bibinfo
  {volume} {414}},\ \bibinfo {pages} {413} (\bibinfo {year}
  {2001})}\BibitemShut {NoStop}%
\bibitem [{\citenamefont {Lago-Rivera}\ \emph {et~al.}(2021)\citenamefont
  {Lago-Rivera}, \citenamefont {Grandi}, \citenamefont {Rakonjac},
  \citenamefont {Seri},\ and\ \citenamefont {de~Riedmatten}}]{Lago2021}%
  \BibitemOpen
  \bibfield  {author} {\bibinfo {author} {\bibfnamefont {D.}~\bibnamefont
  {Lago-Rivera}}, \bibinfo {author} {\bibfnamefont {S.}~\bibnamefont {Grandi}},
  \bibinfo {author} {\bibfnamefont {J.~V.}\ \bibnamefont {Rakonjac}}, \bibinfo
  {author} {\bibfnamefont {A.}~\bibnamefont {Seri}},\ and\ \bibinfo {author}
  {\bibfnamefont {H.}~\bibnamefont {de~Riedmatten}},\ }\bibfield  {title}
  {\bibinfo {title} {Telecom-heralded entanglement between multimode
  solid-state quantum memories},\ }\href
  {https://doi.org/10.1038/s41586-021-03481-8} {\bibfield  {journal} {\bibinfo
  {journal} {Nature}\ }\textbf {\bibinfo {volume} {594}},\ \bibinfo {pages}
  {37} (\bibinfo {year} {2021})}\BibitemShut {NoStop}%
\bibitem [{\citenamefont {Li}\ \emph {et~al.}(2021)\citenamefont {Li},
  \citenamefont {Dou}, \citenamefont {Pang}, \citenamefont {Yang},
  \citenamefont {Zhang}, \citenamefont {Chen}, \citenamefont {Li},
  \citenamefont {Walmsley},\ and\ \citenamefont {Jin}}]{Li2021}%
  \BibitemOpen
  \bibfield  {author} {\bibinfo {author} {\bibfnamefont {H.}~\bibnamefont
  {Li}}, \bibinfo {author} {\bibfnamefont {J.-P.}\ \bibnamefont {Dou}},
  \bibinfo {author} {\bibfnamefont {X.-L.}\ \bibnamefont {Pang}}, \bibinfo
  {author} {\bibfnamefont {T.-H.}\ \bibnamefont {Yang}}, \bibinfo {author}
  {\bibfnamefont {C.-N.}\ \bibnamefont {Zhang}}, \bibinfo {author}
  {\bibfnamefont {Y.}~\bibnamefont {Chen}}, \bibinfo {author} {\bibfnamefont
  {J.-M.}\ \bibnamefont {Li}}, \bibinfo {author} {\bibfnamefont {I.~A.}\
  \bibnamefont {Walmsley}},\ and\ \bibinfo {author} {\bibfnamefont {X.-M.}\
  \bibnamefont {Jin}},\ }\bibfield  {title} {\bibinfo {title} {Heralding
  quantum entanglement between two room-temperature atomic ensembles},\ }\href
  {https://doi.org/10.1364/OPTICA.424599} {\bibfield  {journal} {\bibinfo
  {journal} {Optica}\ }\textbf {\bibinfo {volume} {8}},\ \bibinfo {pages} {925}
  (\bibinfo {year} {2021})}\BibitemShut {NoStop}%
\bibitem [{\citenamefont {Businger}\ \emph {et~al.}(2022)\citenamefont
  {Businger}, \citenamefont {Nicolas}, \citenamefont {Mejia}, \citenamefont
  {Ferrier}, \citenamefont {Goldner},\ and\ \citenamefont
  {Afzelius}}]{Businger2022}%
  \BibitemOpen
  \bibfield  {author} {\bibinfo {author} {\bibfnamefont {M.}~\bibnamefont
  {Businger}}, \bibinfo {author} {\bibfnamefont {L.}~\bibnamefont {Nicolas}},
  \bibinfo {author} {\bibfnamefont {T.~S.}\ \bibnamefont {Mejia}}, \bibinfo
  {author} {\bibfnamefont {A.}~\bibnamefont {Ferrier}}, \bibinfo {author}
  {\bibfnamefont {P.}~\bibnamefont {Goldner}},\ and\ \bibinfo {author}
  {\bibfnamefont {M.}~\bibnamefont {Afzelius}},\ }\bibfield  {title} {\bibinfo
  {title} {Non-classical correlations over 1250 modes between telecom photons
  and 979-nm photons stored in 171yb3+:y2sio5},\ }\href
  {https://doi.org/10.1038/s41467-022-33929-y} {\bibfield  {journal} {\bibinfo
  {journal} {Nature Communications}\ }\textbf {\bibinfo {volume} {13}},\
  \bibinfo {pages} {6438} (\bibinfo {year} {2022})}\BibitemShut {NoStop}%
\bibitem [{\citenamefont {Krovi}\ \emph {et~al.}(2016)\citenamefont {Krovi},
  \citenamefont {Guha}, \citenamefont {Dutton}, \citenamefont {Slater},
  \citenamefont {Simon},\ and\ \citenamefont {Tittel}}]{Krovi2016}%
  \BibitemOpen
  \bibfield  {author} {\bibinfo {author} {\bibfnamefont {H.}~\bibnamefont
  {Krovi}}, \bibinfo {author} {\bibfnamefont {S.}~\bibnamefont {Guha}},
  \bibinfo {author} {\bibfnamefont {Z.}~\bibnamefont {Dutton}}, \bibinfo
  {author} {\bibfnamefont {J.~A.}\ \bibnamefont {Slater}}, \bibinfo {author}
  {\bibfnamefont {C.}~\bibnamefont {Simon}},\ and\ \bibinfo {author}
  {\bibfnamefont {W.}~\bibnamefont {Tittel}},\ }\bibfield  {title} {\bibinfo
  {title} {Practical quantum repeaters with parametric down-conversion
  sources},\ }\href {https://doi.org/10.1007/s00340-015-6297-4} {\bibfield
  {journal} {\bibinfo  {journal} {Applied Physics B}\ }\textbf {\bibinfo
  {volume} {122}},\ \bibinfo {pages} {52} (\bibinfo {year} {2016})}\BibitemShut
  {NoStop}%
\bibitem [{\citenamefont {Yoshida}\ and\ \citenamefont
  {Horikiri}(2024)}]{yoshida2024multiplexed}%
  \BibitemOpen
  \bibfield  {author} {\bibinfo {author} {\bibfnamefont {D.}~\bibnamefont
  {Yoshida}}\ and\ \bibinfo {author} {\bibfnamefont {T.}~\bibnamefont
  {Horikiri}},\ }\href@noop {} {\bibinfo {title} {A multiplexed quantum
  repeater based on single-photon interference with mild stabilization}}
  (\bibinfo {year} {2024}),\ \Eprint {https://arxiv.org/abs/2401.09578}
  {arXiv:2401.09578 [quant-ph]} \BibitemShut {NoStop}%
\bibitem [{\citenamefont {Langenfeld}\ \emph {et~al.}(2021)\citenamefont
  {Langenfeld}, \citenamefont {Thomas}, \citenamefont {Morin},\ and\
  \citenamefont {Rempe}}]{Langenfeld2021}%
  \BibitemOpen
  \bibfield  {author} {\bibinfo {author} {\bibfnamefont {S.}~\bibnamefont
  {Langenfeld}}, \bibinfo {author} {\bibfnamefont {P.}~\bibnamefont {Thomas}},
  \bibinfo {author} {\bibfnamefont {O.}~\bibnamefont {Morin}},\ and\ \bibinfo
  {author} {\bibfnamefont {G.}~\bibnamefont {Rempe}},\ }\bibfield  {title}
  {\bibinfo {title} {Quantum repeater node demonstrating unconditionally secure
  key distribution},\ }\href {https://doi.org/10.1103/PhysRevLett.126.230506}
  {\bibfield  {journal} {\bibinfo  {journal} {Phys. Rev. Lett.}\ }\textbf
  {\bibinfo {volume} {126}},\ \bibinfo {pages} {230506} (\bibinfo {year}
  {2021})}\BibitemShut {NoStop}%
\bibitem [{\citenamefont {Morin}\ \emph {et~al.}(2019)\citenamefont {Morin},
  \citenamefont {K\"orber}, \citenamefont {Langenfeld},\ and\ \citenamefont
  {Rempe}}]{Morin2021}%
  \BibitemOpen
  \bibfield  {author} {\bibinfo {author} {\bibfnamefont {O.}~\bibnamefont
  {Morin}}, \bibinfo {author} {\bibfnamefont {M.}~\bibnamefont {K\"orber}},
  \bibinfo {author} {\bibfnamefont {S.}~\bibnamefont {Langenfeld}},\ and\
  \bibinfo {author} {\bibfnamefont {G.}~\bibnamefont {Rempe}},\ }\bibfield
  {title} {\bibinfo {title} {Deterministic shaping and reshaping of
  single-photon temporal wave functions},\ }\href
  {https://doi.org/10.1103/PhysRevLett.123.133602} {\bibfield  {journal}
  {\bibinfo  {journal} {Phys. Rev. Lett.}\ }\textbf {\bibinfo {volume} {123}},\
  \bibinfo {pages} {133602} (\bibinfo {year} {2019})}\BibitemShut {NoStop}%
\bibitem [{\citenamefont {Knall}\ \emph {et~al.}(2022)\citenamefont {Knall},
  \citenamefont {Knaut}, \citenamefont {Bekenstein}, \citenamefont {Assumpcao},
  \citenamefont {Stroganov}, \citenamefont {Gong}, \citenamefont {Huan},
  \citenamefont {Stas}, \citenamefont {Machielse}, \citenamefont {Chalupnik},
  \citenamefont {Levonian}, \citenamefont {Suleymanzade}, \citenamefont
  {Riedinger}, \citenamefont {Park}, \citenamefont {Lon\ifmmode~\check{c}\else
  \v{c}\fi{}ar}, \citenamefont {Bhaskar},\ and\ \citenamefont
  {Lukin}}]{Knall2022}%
  \BibitemOpen
  \bibfield  {author} {\bibinfo {author} {\bibfnamefont {E.~N.}\ \bibnamefont
  {Knall}}, \bibinfo {author} {\bibfnamefont {C.~M.}\ \bibnamefont {Knaut}},
  \bibinfo {author} {\bibfnamefont {R.}~\bibnamefont {Bekenstein}}, \bibinfo
  {author} {\bibfnamefont {D.~R.}\ \bibnamefont {Assumpcao}}, \bibinfo {author}
  {\bibfnamefont {P.~L.}\ \bibnamefont {Stroganov}}, \bibinfo {author}
  {\bibfnamefont {W.}~\bibnamefont {Gong}}, \bibinfo {author} {\bibfnamefont
  {Y.~Q.}\ \bibnamefont {Huan}}, \bibinfo {author} {\bibfnamefont {P.-J.}\
  \bibnamefont {Stas}}, \bibinfo {author} {\bibfnamefont {B.}~\bibnamefont
  {Machielse}}, \bibinfo {author} {\bibfnamefont {M.}~\bibnamefont
  {Chalupnik}}, \bibinfo {author} {\bibfnamefont {D.}~\bibnamefont {Levonian}},
  \bibinfo {author} {\bibfnamefont {A.}~\bibnamefont {Suleymanzade}}, \bibinfo
  {author} {\bibfnamefont {R.}~\bibnamefont {Riedinger}}, \bibinfo {author}
  {\bibfnamefont {H.}~\bibnamefont {Park}}, \bibinfo {author} {\bibfnamefont
  {M.}~\bibnamefont {Lon\ifmmode~\check{c}\else \v{c}\fi{}ar}}, \bibinfo
  {author} {\bibfnamefont {M.~K.}\ \bibnamefont {Bhaskar}},\ and\ \bibinfo
  {author} {\bibfnamefont {M.~D.}\ \bibnamefont {Lukin}},\ }\bibfield  {title}
  {\bibinfo {title} {Efficient source of shaped single photons based on an
  integrated diamond nanophotonic system},\ }\href
  {https://doi.org/10.1103/PhysRevLett.129.053603} {\bibfield  {journal}
  {\bibinfo  {journal} {Phys. Rev. Lett.}\ }\textbf {\bibinfo {volume} {129}},\
  \bibinfo {pages} {053603} (\bibinfo {year} {2022})}\BibitemShut {NoStop}%
\bibitem [{\citenamefont {Raha}\ \emph {et~al.}(2020)\citenamefont {Raha},
  \citenamefont {Chen}, \citenamefont {Phenicie}, \citenamefont {Ourari},
  \citenamefont {Dibos},\ and\ \citenamefont {Thompson}}]{Raha2020}%
  \BibitemOpen
  \bibfield  {author} {\bibinfo {author} {\bibfnamefont {M.}~\bibnamefont
  {Raha}}, \bibinfo {author} {\bibfnamefont {S.}~\bibnamefont {Chen}}, \bibinfo
  {author} {\bibfnamefont {C.~M.}\ \bibnamefont {Phenicie}}, \bibinfo {author}
  {\bibfnamefont {S.}~\bibnamefont {Ourari}}, \bibinfo {author} {\bibfnamefont
  {A.~M.}\ \bibnamefont {Dibos}},\ and\ \bibinfo {author} {\bibfnamefont
  {J.~D.}\ \bibnamefont {Thompson}},\ }\bibfield  {title} {\bibinfo {title}
  {Optical quantum nondemolition measurement of a single rare earth ion
  qubit},\ }\href {https://doi.org/10.1038/s41467-020-15138-7} {\bibfield
  {journal} {\bibinfo  {journal} {Nature Communications}\ }\textbf {\bibinfo
  {volume} {11}},\ \bibinfo {pages} {1605} (\bibinfo {year}
  {2020})}\BibitemShut {NoStop}%
\bibitem [{\citenamefont {Kalb}\ \emph {et~al.}(2017)\citenamefont {Kalb},
  \citenamefont {Reiserer}, \citenamefont {Humphreys}, \citenamefont
  {Bakermans}, \citenamefont {Kamerling}, \citenamefont {Nickerson},
  \citenamefont {Benjamin}, \citenamefont {Twitchen}, \citenamefont {Markham},\
  and\ \citenamefont {Hanson}}]{Kalb2017}%
  \BibitemOpen
  \bibfield  {author} {\bibinfo {author} {\bibfnamefont {N.}~\bibnamefont
  {Kalb}}, \bibinfo {author} {\bibfnamefont {A.~A.}\ \bibnamefont {Reiserer}},
  \bibinfo {author} {\bibfnamefont {P.~C.}\ \bibnamefont {Humphreys}}, \bibinfo
  {author} {\bibfnamefont {J.~J.~W.}\ \bibnamefont {Bakermans}}, \bibinfo
  {author} {\bibfnamefont {S.~J.}\ \bibnamefont {Kamerling}}, \bibinfo {author}
  {\bibfnamefont {N.~H.}\ \bibnamefont {Nickerson}}, \bibinfo {author}
  {\bibfnamefont {S.~C.}\ \bibnamefont {Benjamin}}, \bibinfo {author}
  {\bibfnamefont {D.~J.}\ \bibnamefont {Twitchen}}, \bibinfo {author}
  {\bibfnamefont {M.}~\bibnamefont {Markham}},\ and\ \bibinfo {author}
  {\bibfnamefont {R.}~\bibnamefont {Hanson}},\ }\bibfield  {title} {\bibinfo
  {title} {Entanglement distillation between solid-state quantum network
  nodes},\ }\href {https://doi.org/10.1126/science.aan0070} {\bibfield
  {journal} {\bibinfo  {journal} {Science}\ }\textbf {\bibinfo {volume}
  {356}},\ \bibinfo {pages} {928} (\bibinfo {year} {2017})}\BibitemShut
  {NoStop}%
\bibitem [{\citenamefont {Tanzilli}\ \emph {et~al.}(2005)\citenamefont
  {Tanzilli}, \citenamefont {Tittel}, \citenamefont {Halder}, \citenamefont
  {Alibard}, \citenamefont {Baldi}, \citenamefont {Gisin},\ and\ \citenamefont
  {Zbinden}}]{Tanzilli2005}%
  \BibitemOpen
  \bibfield  {author} {\bibinfo {author} {\bibfnamefont {S.}~\bibnamefont
  {Tanzilli}}, \bibinfo {author} {\bibfnamefont {W.}~\bibnamefont {Tittel}},
  \bibinfo {author} {\bibfnamefont {M.}~\bibnamefont {Halder}}, \bibinfo
  {author} {\bibfnamefont {O.}~\bibnamefont {Alibard}}, \bibinfo {author}
  {\bibfnamefont {P.}~\bibnamefont {Baldi}}, \bibinfo {author} {\bibfnamefont
  {N.}~\bibnamefont {Gisin}},\ and\ \bibinfo {author} {\bibfnamefont
  {H.}~\bibnamefont {Zbinden}},\ }\bibfield  {title} {\bibinfo {title} {A
  photonic quantum information interface},\ }\href
  {https://pubmed.ncbi.nlm.nih.gov/16136138/} {\bibfield  {journal} {\bibinfo
  {journal} {Nature}\ }\textbf {\bibinfo {volume} {437}},\ \bibinfo {pages}
  {116} (\bibinfo {year} {2005})}\BibitemShut {NoStop}%
\bibitem [{\citenamefont {Reiserer}\ \emph {et~al.}(2014)\citenamefont
  {Reiserer}, \citenamefont {Kalb}, \citenamefont {Rempe},\ and\ \citenamefont
  {Ritter}}]{Reiserer2014}%
  \BibitemOpen
  \bibfield  {author} {\bibinfo {author} {\bibfnamefont {A.}~\bibnamefont
  {Reiserer}}, \bibinfo {author} {\bibfnamefont {N.}~\bibnamefont {Kalb}},
  \bibinfo {author} {\bibfnamefont {G.}~\bibnamefont {Rempe}},\ and\ \bibinfo
  {author} {\bibfnamefont {S.}~\bibnamefont {Ritter}},\ }\bibfield  {title}
  {\bibinfo {title} {A quantum gate between a flying optical photon and a
  single trapped atom},\ }\href {https://doi.org/10.1038/nature13177}
  {\bibfield  {journal} {\bibinfo  {journal} {Nature}\ }\textbf {\bibinfo
  {volume} {508}},\ \bibinfo {pages} {237} (\bibinfo {year}
  {2014})}\BibitemShut {NoStop}%
\bibitem [{\citenamefont {Kalb}\ \emph {et~al.}(2015)\citenamefont {Kalb},
  \citenamefont {Reiserer}, \citenamefont {Ritter},\ and\ \citenamefont
  {Rempe}}]{Kalb2015}%
  \BibitemOpen
  \bibfield  {author} {\bibinfo {author} {\bibfnamefont {N.}~\bibnamefont
  {Kalb}}, \bibinfo {author} {\bibfnamefont {A.}~\bibnamefont {Reiserer}},
  \bibinfo {author} {\bibfnamefont {S.}~\bibnamefont {Ritter}},\ and\ \bibinfo
  {author} {\bibfnamefont {G.}~\bibnamefont {Rempe}},\ }\bibfield  {title}
  {\bibinfo {title} {Heralded storage of a photonic quantum bit in a single
  atom},\ }\href {https://doi.org/10.1103/PhysRevLett.114.220501} {\bibfield
  {journal} {\bibinfo  {journal} {Phys. Rev. Lett.}\ }\textbf {\bibinfo
  {volume} {114}},\ \bibinfo {pages} {220501} (\bibinfo {year}
  {2015})}\BibitemShut {NoStop}%
\bibitem [{\citenamefont {Gu}(2024)}]{codeCollection}%
  \BibitemOpen
  \bibfield  {author} {\bibinfo {author} {\bibfnamefont {F.}~\bibnamefont
  {Gu}},\ }\href
  {https://doi.org/10.4121/589448ac-1be7-43a7-988d-c4c1122eb4e8.v1} {\bibinfo
  {title} {Codes underlying the publication: "hybrid quantum repeaters with
  ensemble-based quantum memories and single-spin photon transducers"}}
  (\bibinfo {year} {2024})\BibitemShut {NoStop}%
\bibitem [{\citenamefont {Burgess}\ \emph {et~al.}(2009)\citenamefont
  {Burgess}, \citenamefont {Loncar}, \citenamefont {McCutcheon},\ and\
  \citenamefont {Zhang}}]{Zhang2009}%
  \BibitemOpen
  \bibfield  {author} {\bibinfo {author} {\bibfnamefont {I.~B.}\ \bibnamefont
  {Burgess}}, \bibinfo {author} {\bibfnamefont {M.}~\bibnamefont {Loncar}},
  \bibinfo {author} {\bibfnamefont {M.~W.}\ \bibnamefont {McCutcheon}},\ and\
  \bibinfo {author} {\bibfnamefont {Y.}~\bibnamefont {Zhang}},\ }\bibfield
  {title} {\bibinfo {title} {Ultra-high-<i>q</i> te/tm dual-polarized photonic
  crystal nanocavities},\ }\href {https://doi.org/10.1364/OL.34.002694}
  {\bibfield  {journal} {\bibinfo  {journal} {Optics Letters, Vol. 34, Issue
  17, pp. 2694-2696}\ }\textbf {\bibinfo {volume} {34}},\ \bibinfo {pages}
  {2694} (\bibinfo {year} {2009})}\BibitemShut {NoStop}%
\bibitem [{\citenamefont {Vučković}\ \emph {et~al.}(2011)\citenamefont
  {Vučković}, \citenamefont {Rivoire},\ and\ \citenamefont
  {Buckley}}]{Rivoire2011}%
  \BibitemOpen
  \bibfield  {author} {\bibinfo {author} {\bibfnamefont {J.}~\bibnamefont
  {Vučković}}, \bibinfo {author} {\bibfnamefont {K.}~\bibnamefont
  {Rivoire}},\ and\ \bibinfo {author} {\bibfnamefont {S.}~\bibnamefont
  {Buckley}},\ }\bibfield  {title} {\bibinfo {title} {Multiply resonant
  photonic crystal nanocavities for nonlinear frequency conversion},\ }\href
  {https://doi.org/10.1364/OE.19.022198} {\bibfield  {journal} {\bibinfo
  {journal} {Optics Express, Vol. 19, Issue 22, pp. 22198-22207}\ }\textbf
  {\bibinfo {volume} {19}},\ \bibinfo {pages} {22198} (\bibinfo {year}
  {2011})}\BibitemShut {NoStop}%
\bibitem [{\citenamefont {Menon}\ \emph {et~al.}(2023)\citenamefont {Menon},
  \citenamefont {Glachman}, \citenamefont {Pompili}, \citenamefont {Dibos},\
  and\ \citenamefont {Bernien}}]{Menon2023}%
  \BibitemOpen
  \bibfield  {author} {\bibinfo {author} {\bibfnamefont {S.~G.}\ \bibnamefont
  {Menon}}, \bibinfo {author} {\bibfnamefont {N.}~\bibnamefont {Glachman}},
  \bibinfo {author} {\bibfnamefont {M.}~\bibnamefont {Pompili}}, \bibinfo
  {author} {\bibfnamefont {A.}~\bibnamefont {Dibos}},\ and\ \bibinfo {author}
  {\bibfnamefont {H.}~\bibnamefont {Bernien}},\ }\bibfield  {title} {\bibinfo
  {title} {An integrated atom array-nanophotonic chip platform with
  background-free imaging},\ }\href {https://arxiv.org/abs/2311.02153}
  {\bibfield  {journal} {\bibinfo  {journal} {arXiv:2311.02153}\ } (\bibinfo
  {year} {2023})}\BibitemShut {NoStop}%
\bibitem [{\citenamefont {Kim}\ \emph {et~al.}(2019)\citenamefont {Kim},
  \citenamefont {Chang}, \citenamefont {Fields}, \citenamefont {Chen},\ and\
  \citenamefont {Hung}}]{Kim2019}%
  \BibitemOpen
  \bibfield  {author} {\bibinfo {author} {\bibfnamefont {M.~E.}\ \bibnamefont
  {Kim}}, \bibinfo {author} {\bibfnamefont {T.-H.}\ \bibnamefont {Chang}},
  \bibinfo {author} {\bibfnamefont {B.~M.}\ \bibnamefont {Fields}}, \bibinfo
  {author} {\bibfnamefont {C.-A.}\ \bibnamefont {Chen}},\ and\ \bibinfo
  {author} {\bibfnamefont {C.-L.}\ \bibnamefont {Hung}},\ }\bibfield  {title}
  {\bibinfo {title} {Trapping single atoms on a nanophotonic circuit with
  configurable tweezer lattices},\ }\href
  {https://doi.org/10.1038/s41467-019-09635-7} {\bibfield  {journal} {\bibinfo
  {journal} {Nature Communications 2019 10:1}\ }\textbf {\bibinfo {volume}
  {10}},\ \bibinfo {pages} {1} (\bibinfo {year} {2019})}\BibitemShut {NoStop}%
\bibitem [{\citenamefont {Thompson}\ \emph
  {et~al.}(2013{\natexlab{a}})\citenamefont {Thompson}, \citenamefont {Tiecke},
  \citenamefont {de~Leon}, \citenamefont {Feist}, \citenamefont {Akimov},
  \citenamefont {Gullans}, \citenamefont {Zibrov}, \citenamefont {Vuletić},\
  and\ \citenamefont {Lukin}}]{Thompson2013}%
  \BibitemOpen
  \bibfield  {author} {\bibinfo {author} {\bibfnamefont {J.~D.}\ \bibnamefont
  {Thompson}}, \bibinfo {author} {\bibfnamefont {T.~G.}\ \bibnamefont
  {Tiecke}}, \bibinfo {author} {\bibfnamefont {N.~P.}\ \bibnamefont {de~Leon}},
  \bibinfo {author} {\bibfnamefont {J.}~\bibnamefont {Feist}}, \bibinfo
  {author} {\bibfnamefont {A.~V.}\ \bibnamefont {Akimov}}, \bibinfo {author}
  {\bibfnamefont {M.}~\bibnamefont {Gullans}}, \bibinfo {author} {\bibfnamefont
  {A.~S.}\ \bibnamefont {Zibrov}}, \bibinfo {author} {\bibfnamefont
  {V.}~\bibnamefont {Vuletić}},\ and\ \bibinfo {author} {\bibfnamefont
  {M.~D.}\ \bibnamefont {Lukin}},\ }\bibfield  {title} {\bibinfo {title}
  {Coupling a single trapped atom to a nanoscale optical cavity},\ }\href
  {https://doi.org/10.1126/SCIENCE.1237125} {\bibfield  {journal} {\bibinfo
  {journal} {Science}\ }\textbf {\bibinfo {volume} {340}},\ \bibinfo {pages}
  {1202} (\bibinfo {year} {2013}{\natexlab{a}})}\BibitemShut {NoStop}%
\bibitem [{\citenamefont {Samutpraphoot}\ \emph {et~al.}(2020)\citenamefont
  {Samutpraphoot}, \citenamefont {Đorđević}, \citenamefont {Ocola},
  \citenamefont {Bernien}, \citenamefont {Senko}, \citenamefont {Vuletić},\
  and\ \citenamefont {Lukin}}]{Samutpraphoot2020}%
  \BibitemOpen
  \bibfield  {author} {\bibinfo {author} {\bibfnamefont {P.}~\bibnamefont
  {Samutpraphoot}}, \bibinfo {author} {\bibfnamefont {T.}~\bibnamefont
  {Đorđević}}, \bibinfo {author} {\bibfnamefont {P.~L.}\ \bibnamefont
  {Ocola}}, \bibinfo {author} {\bibfnamefont {H.}~\bibnamefont {Bernien}},
  \bibinfo {author} {\bibfnamefont {C.}~\bibnamefont {Senko}}, \bibinfo
  {author} {\bibfnamefont {V.}~\bibnamefont {Vuletić}},\ and\ \bibinfo
  {author} {\bibfnamefont {M.~D.}\ \bibnamefont {Lukin}},\ }\bibfield  {title}
  {\bibinfo {title} {Strong coupling of two individually controlled atoms via a
  nanophotonic cavity},\ }\href
  {https://doi.org/10.1103/PhysRevLett.124.063602} {\bibfield  {journal}
  {\bibinfo  {journal} {Physical Review Letters}\ }\textbf {\bibinfo {volume}
  {124}},\ \bibinfo {pages} {063602} (\bibinfo {year} {2020})}\BibitemShut
  {NoStop}%
\bibitem [{\citenamefont {Thompson}\ \emph
  {et~al.}(2013{\natexlab{b}})\citenamefont {Thompson}, \citenamefont {Tiecke},
  \citenamefont {Zibrov}, \citenamefont {Vuleti{\'c}},\ and\ \citenamefont
  {Lukin}}]{Thompson_cooling2013}%
  \BibitemOpen
  \bibfield  {author} {\bibinfo {author} {\bibfnamefont {J.~D.}\ \bibnamefont
  {Thompson}}, \bibinfo {author} {\bibfnamefont {T.}~\bibnamefont {Tiecke}},
  \bibinfo {author} {\bibfnamefont {A.~S.}\ \bibnamefont {Zibrov}}, \bibinfo
  {author} {\bibfnamefont {V.}~\bibnamefont {Vuleti{\'c}}},\ and\ \bibinfo
  {author} {\bibfnamefont {M.~D.}\ \bibnamefont {Lukin}},\ }\bibfield  {title}
  {\bibinfo {title} {Coherence and raman sideband cooling of a single atom in
  an optical tweezer},\ }\href
  {https://journals.aps.org/prl/abstract/10.1103/PhysRevLett.110.133001}
  {\bibfield  {journal} {\bibinfo  {journal} {Physical review letters}\
  }\textbf {\bibinfo {volume} {110}},\ \bibinfo {pages} {133001} (\bibinfo
  {year} {2013}{\natexlab{b}})}\BibitemShut {NoStop}%
\bibitem [{\citenamefont {Kaufman}\ \emph {et~al.}(2012)\citenamefont
  {Kaufman}, \citenamefont {Lester},\ and\ \citenamefont
  {Regal}}]{Kaufman2012}%
  \BibitemOpen
  \bibfield  {author} {\bibinfo {author} {\bibfnamefont {A.~M.}\ \bibnamefont
  {Kaufman}}, \bibinfo {author} {\bibfnamefont {B.~J.}\ \bibnamefont
  {Lester}},\ and\ \bibinfo {author} {\bibfnamefont {C.~A.}\ \bibnamefont
  {Regal}},\ }\bibfield  {title} {\bibinfo {title} {Cooling a single atom in an
  optical tweezer to its quantum ground state},\ }\href
  {https://journals.aps.org/prx/abstract/10.1103/PhysRevX.2.041014} {\bibfield
  {journal} {\bibinfo  {journal} {Physical Review X}\ }\textbf {\bibinfo
  {volume} {2}},\ \bibinfo {pages} {041014} (\bibinfo {year}
  {2012})}\BibitemShut {NoStop}%
\bibitem [{\citenamefont {Lvovsky}\ \emph {et~al.}(2009)\citenamefont
  {Lvovsky}, \citenamefont {Sanders},\ and\ \citenamefont {Tittel}}]{QM1}%
  \BibitemOpen
  \bibfield  {author} {\bibinfo {author} {\bibfnamefont {A.~I.}\ \bibnamefont
  {Lvovsky}}, \bibinfo {author} {\bibfnamefont {B.~C.}\ \bibnamefont
  {Sanders}},\ and\ \bibinfo {author} {\bibfnamefont {W.}~\bibnamefont
  {Tittel}},\ }\bibfield  {title} {\bibinfo {title} {Optical quantum memory},\
  }\href {https://doi.org/10.1038/nphoton.2009.231} {\bibfield  {journal}
  {\bibinfo  {journal} {Nature Photonics}\ }\textbf {\bibinfo {volume} {3}},\
  \bibinfo {pages} {706} (\bibinfo {year} {2009})}\BibitemShut {NoStop}%
\bibitem [{\citenamefont {Jing}\ and\ \citenamefont
  {Bao}(2023)}]{jing2023ensemble}%
  \BibitemOpen
  \bibfield  {author} {\bibinfo {author} {\bibfnamefont {B.}~\bibnamefont
  {Jing}}\ and\ \bibinfo {author} {\bibfnamefont {X.-H.}\ \bibnamefont {Bao}},\
  }\bibfield  {title} {\bibinfo {title} {Ensemble-based quantum memory:
  Principle, advance, and application},\ }\href@noop {} {\bibfield  {journal}
  {\bibinfo  {journal} {Photonic Quantum Technologies: Science and
  Applications}\ }\textbf {\bibinfo {volume} {2}},\ \bibinfo {pages} {433}
  (\bibinfo {year} {2023})}\BibitemShut {NoStop}%
\bibitem [{\citenamefont {Ma}\ \emph {et~al.}(2017)\citenamefont {Ma},
  \citenamefont {Slattery},\ and\ \citenamefont {Tang}}]{ma2017optical}%
  \BibitemOpen
  \bibfield  {author} {\bibinfo {author} {\bibfnamefont {L.}~\bibnamefont
  {Ma}}, \bibinfo {author} {\bibfnamefont {O.}~\bibnamefont {Slattery}},\ and\
  \bibinfo {author} {\bibfnamefont {X.}~\bibnamefont {Tang}},\ }\bibfield
  {title} {\bibinfo {title} {Optical quantum memory based on
  electromagnetically induced transparency},\ }\href@noop {} {\bibfield
  {journal} {\bibinfo  {journal} {Journal of Optics}\ }\textbf {\bibinfo
  {volume} {19}},\ \bibinfo {pages} {043001} (\bibinfo {year}
  {2017})}\BibitemShut {NoStop}%
\bibitem [{\citenamefont {Bussières}\ \emph {et~al.}(2013)\citenamefont
  {Bussières}, \citenamefont {Sangouard}, \citenamefont {Afzelius},
  \citenamefont {de~Riedmatten}, \citenamefont {Simon},\ and\ \citenamefont
  {Tittel}}]{QM2_echo}%
  \BibitemOpen
  \bibfield  {author} {\bibinfo {author} {\bibfnamefont {F.}~\bibnamefont
  {Bussières}}, \bibinfo {author} {\bibfnamefont {N.}~\bibnamefont
  {Sangouard}}, \bibinfo {author} {\bibfnamefont {M.}~\bibnamefont {Afzelius}},
  \bibinfo {author} {\bibfnamefont {H.}~\bibnamefont {de~Riedmatten}}, \bibinfo
  {author} {\bibfnamefont {C.}~\bibnamefont {Simon}},\ and\ \bibinfo {author}
  {\bibfnamefont {W.}~\bibnamefont {Tittel}},\ }\bibfield  {title} {\bibinfo
  {title} {Prospective applications of optical quantum memories},\ }\href
  {https://doi.org/10.1080/09500340.2013.856482} {\bibfield  {journal}
  {\bibinfo  {journal} {Journal of Modern Optics}\ }\textbf {\bibinfo {volume}
  {60}},\ \bibinfo {pages} {1519} (\bibinfo {year} {2013})}\BibitemShut
  {NoStop}%
\bibitem [{\citenamefont {Afzelius}\ \emph {et~al.}(2009)\citenamefont
  {Afzelius}, \citenamefont {Simon}, \citenamefont {de~Riedmatten},\ and\
  \citenamefont {Gisin}}]{AFC}%
  \BibitemOpen
  \bibfield  {author} {\bibinfo {author} {\bibfnamefont {M.}~\bibnamefont
  {Afzelius}}, \bibinfo {author} {\bibfnamefont {C.}~\bibnamefont {Simon}},
  \bibinfo {author} {\bibfnamefont {H.}~\bibnamefont {de~Riedmatten}},\ and\
  \bibinfo {author} {\bibfnamefont {N.}~\bibnamefont {Gisin}},\ }\bibfield
  {title} {\bibinfo {title} {Multimode quantum memory based on atomic frequency
  combs},\ }\href {https://doi.org/10.1103/PhysRevA.79.052329} {\bibfield
  {journal} {\bibinfo  {journal} {Phys. Rev. A}\ }\textbf {\bibinfo {volume}
  {79}},\ \bibinfo {pages} {052329} (\bibinfo {year} {2009})}\BibitemShut
  {NoStop}%
\bibitem [{\citenamefont {Jobez}\ \emph {et~al.}(2016)\citenamefont {Jobez},
  \citenamefont {Timoney}, \citenamefont {Laplane}, \citenamefont {Etesse},
  \citenamefont {Ferrier}, \citenamefont {Goldner}, \citenamefont {Gisin},\
  and\ \citenamefont {Afzelius}}]{jobez2016}%
  \BibitemOpen
  \bibfield  {author} {\bibinfo {author} {\bibfnamefont {P.}~\bibnamefont
  {Jobez}}, \bibinfo {author} {\bibfnamefont {N.}~\bibnamefont {Timoney}},
  \bibinfo {author} {\bibfnamefont {C.}~\bibnamefont {Laplane}}, \bibinfo
  {author} {\bibfnamefont {J.}~\bibnamefont {Etesse}}, \bibinfo {author}
  {\bibfnamefont {A.}~\bibnamefont {Ferrier}}, \bibinfo {author} {\bibfnamefont
  {P.}~\bibnamefont {Goldner}}, \bibinfo {author} {\bibfnamefont
  {N.}~\bibnamefont {Gisin}},\ and\ \bibinfo {author} {\bibfnamefont
  {M.}~\bibnamefont {Afzelius}},\ }\bibfield  {title} {\bibinfo {title}
  {Towards highly multimode optical quantum memory for quantum repeaters},\
  }\href {https://doi.org/10.1103/PhysRevA.93.032327} {\bibfield  {journal}
  {\bibinfo  {journal} {Phys. Rev. A}\ }\textbf {\bibinfo {volume} {93}},\
  \bibinfo {pages} {032327} (\bibinfo {year} {2016})}\BibitemShut {NoStop}%
\bibitem [{\citenamefont {Bashkansky}\ \emph {et~al.}(2012)\citenamefont
  {Bashkansky}, \citenamefont {Fatemi},\ and\ \citenamefont
  {Vurgaftman}}]{hot_Rb}%
  \BibitemOpen
  \bibfield  {author} {\bibinfo {author} {\bibfnamefont {M.}~\bibnamefont
  {Bashkansky}}, \bibinfo {author} {\bibfnamefont {F.~K.}\ \bibnamefont
  {Fatemi}},\ and\ \bibinfo {author} {\bibfnamefont {I.}~\bibnamefont
  {Vurgaftman}},\ }\bibfield  {title} {\bibinfo {title} {Quantum memory in warm
  rubidium vapor with buffer gas},\ }\href@noop {} {\bibfield  {journal}
  {\bibinfo  {journal} {Optics Letters}\ }\textbf {\bibinfo {volume} {37}},\
  \bibinfo {pages} {142} (\bibinfo {year} {2012})}\BibitemShut {NoStop}%
\bibitem [{\citenamefont {Vernaz-Gris}\ \emph {et~al.}(2018)\citenamefont
  {Vernaz-Gris}, \citenamefont {Huang}, \citenamefont {Cao}, \citenamefont
  {Sheremet},\ and\ \citenamefont {Laurat}}]{Efficient_storage_1}%
  \BibitemOpen
  \bibfield  {author} {\bibinfo {author} {\bibfnamefont {P.}~\bibnamefont
  {Vernaz-Gris}}, \bibinfo {author} {\bibfnamefont {K.}~\bibnamefont {Huang}},
  \bibinfo {author} {\bibfnamefont {M.}~\bibnamefont {Cao}}, \bibinfo {author}
  {\bibfnamefont {A.~S.}\ \bibnamefont {Sheremet}},\ and\ \bibinfo {author}
  {\bibfnamefont {J.}~\bibnamefont {Laurat}},\ }\bibfield  {title} {\bibinfo
  {title} {Highly-efficient quantum memory for polarization qubits in a
  spatially-multiplexed cold atomic ensemble},\ }\href@noop {} {\bibfield
  {journal} {\bibinfo  {journal} {Nature communications}\ }\textbf {\bibinfo
  {volume} {9}},\ \bibinfo {pages} {1} (\bibinfo {year} {2018})}\BibitemShut
  {NoStop}%
\bibitem [{\citenamefont {Cho}\ \emph {et~al.}(2016)\citenamefont {Cho},
  \citenamefont {Campbell}, \citenamefont {Everett}, \citenamefont {Bernu},
  \citenamefont {Higginbottom}, \citenamefont {Cao}, \citenamefont {Geng},
  \citenamefont {Robins}, \citenamefont {Lam},\ and\ \citenamefont
  {Buchler}}]{Efficient_storage_2}%
  \BibitemOpen
  \bibfield  {author} {\bibinfo {author} {\bibfnamefont {Y.-W.}\ \bibnamefont
  {Cho}}, \bibinfo {author} {\bibfnamefont {G.}~\bibnamefont {Campbell}},
  \bibinfo {author} {\bibfnamefont {J.}~\bibnamefont {Everett}}, \bibinfo
  {author} {\bibfnamefont {J.}~\bibnamefont {Bernu}}, \bibinfo {author}
  {\bibfnamefont {D.}~\bibnamefont {Higginbottom}}, \bibinfo {author}
  {\bibfnamefont {M.}~\bibnamefont {Cao}}, \bibinfo {author} {\bibfnamefont
  {J.}~\bibnamefont {Geng}}, \bibinfo {author} {\bibfnamefont {N.}~\bibnamefont
  {Robins}}, \bibinfo {author} {\bibfnamefont {P.}~\bibnamefont {Lam}},\ and\
  \bibinfo {author} {\bibfnamefont {B.}~\bibnamefont {Buchler}},\ }\bibfield
  {title} {\bibinfo {title} {Highly efficient optical quantum memory with long
  coherence time in cold atoms},\ }\href@noop {} {\bibfield  {journal}
  {\bibinfo  {journal} {Optica}\ }\textbf {\bibinfo {volume} {3}},\ \bibinfo
  {pages} {100} (\bibinfo {year} {2016})}\BibitemShut {NoStop}%
\bibitem [{\citenamefont {Hosseini}\ \emph {et~al.}(2011)\citenamefont
  {Hosseini}, \citenamefont {Sparkes}, \citenamefont {Campbell}, \citenamefont
  {Lam},\ and\ \citenamefont {Buchler}}]{Efficient_storage_3}%
  \BibitemOpen
  \bibfield  {author} {\bibinfo {author} {\bibfnamefont {M.}~\bibnamefont
  {Hosseini}}, \bibinfo {author} {\bibfnamefont {B.~M.}\ \bibnamefont
  {Sparkes}}, \bibinfo {author} {\bibfnamefont {G.}~\bibnamefont {Campbell}},
  \bibinfo {author} {\bibfnamefont {P.~K.}\ \bibnamefont {Lam}},\ and\ \bibinfo
  {author} {\bibfnamefont {B.~C.}\ \bibnamefont {Buchler}},\ }\bibfield
  {title} {\bibinfo {title} {High efficiency coherent optical memory with warm
  rubidium vapour},\ }\href@noop {} {\bibfield  {journal} {\bibinfo  {journal}
  {Nature communications}\ }\textbf {\bibinfo {volume} {2}},\ \bibinfo {pages}
  {1} (\bibinfo {year} {2011})}\BibitemShut {NoStop}%
\bibitem [{\citenamefont {Guo}\ \emph {et~al.}(2019)\citenamefont {Guo},
  \citenamefont {Feng}, \citenamefont {Yang}, \citenamefont {Yu}, \citenamefont
  {Chen}, \citenamefont {Yuan},\ and\ \citenamefont {Zhang}}]{Rb_Raman}%
  \BibitemOpen
  \bibfield  {author} {\bibinfo {author} {\bibfnamefont {J.}~\bibnamefont
  {Guo}}, \bibinfo {author} {\bibfnamefont {X.}~\bibnamefont {Feng}}, \bibinfo
  {author} {\bibfnamefont {P.}~\bibnamefont {Yang}}, \bibinfo {author}
  {\bibfnamefont {Z.}~\bibnamefont {Yu}}, \bibinfo {author} {\bibfnamefont
  {L.}~\bibnamefont {Chen}}, \bibinfo {author} {\bibfnamefont {C.-H.}\
  \bibnamefont {Yuan}},\ and\ \bibinfo {author} {\bibfnamefont
  {W.}~\bibnamefont {Zhang}},\ }\bibfield  {title} {\bibinfo {title}
  {High-performance raman quantum memory with optimal control in room
  temperature atoms},\ }\href@noop {} {\bibfield  {journal} {\bibinfo
  {journal} {Nature communications}\ }\textbf {\bibinfo {volume} {10}},\
  \bibinfo {pages} {1} (\bibinfo {year} {2019})}\BibitemShut {NoStop}%
\bibitem [{\citenamefont {Tittel}\ \emph {et~al.}(2010)\citenamefont {Tittel},
  \citenamefont {Afzelius}, \citenamefont {Chaneliére}, \citenamefont {Cone},
  \citenamefont {Kröll}, \citenamefont {Moiseev},\ and\ \citenamefont
  {Sellars}}]{REI1}%
  \BibitemOpen
  \bibfield  {author} {\bibinfo {author} {\bibfnamefont {W.}~\bibnamefont
  {Tittel}}, \bibinfo {author} {\bibfnamefont {M.}~\bibnamefont {Afzelius}},
  \bibinfo {author} {\bibfnamefont {T.}~\bibnamefont {Chaneliére}}, \bibinfo
  {author} {\bibfnamefont {R.}~\bibnamefont {Cone}}, \bibinfo {author}
  {\bibfnamefont {S.}~\bibnamefont {Kröll}}, \bibinfo {author} {\bibfnamefont
  {S.}~\bibnamefont {Moiseev}},\ and\ \bibinfo {author} {\bibfnamefont
  {M.}~\bibnamefont {Sellars}},\ }\bibfield  {title} {\bibinfo {title}
  {Photon-echo quantum memory in solid state systems},\ }\href
  {https://doi.org/https://doi.org/10.1002/lpor.200810056} {\bibfield
  {journal} {\bibinfo  {journal} {Laser \& Photonics Reviews}\ }\textbf
  {\bibinfo {volume} {4}},\ \bibinfo {pages} {244} (\bibinfo {year}
  {2010})}\BibitemShut {NoStop}%
\bibitem [{\citenamefont {Thiel}\ \emph {et~al.}(2011)\citenamefont {Thiel},
  \citenamefont {B{\"o}ttger},\ and\ \citenamefont {Cone}}]{REI2}%
  \BibitemOpen
  \bibfield  {author} {\bibinfo {author} {\bibfnamefont {C.}~\bibnamefont
  {Thiel}}, \bibinfo {author} {\bibfnamefont {T.}~\bibnamefont {B{\"o}ttger}},\
  and\ \bibinfo {author} {\bibfnamefont {R.}~\bibnamefont {Cone}},\ }\bibfield
  {title} {\bibinfo {title} {Rare-earth-doped materials for applications in
  quantum information storage and signal processing},\ }\href@noop {}
  {\bibfield  {journal} {\bibinfo  {journal} {Journal of luminescence}\
  }\textbf {\bibinfo {volume} {131}},\ \bibinfo {pages} {353} (\bibinfo {year}
  {2011})}\BibitemShut {NoStop}%
\bibitem [{\citenamefont {Zhong}\ and\ \citenamefont {Goldner}(2019)}]{REI3}%
  \BibitemOpen
  \bibfield  {author} {\bibinfo {author} {\bibfnamefont {T.}~\bibnamefont
  {Zhong}}\ and\ \bibinfo {author} {\bibfnamefont {P.}~\bibnamefont
  {Goldner}},\ }\bibfield  {title} {\bibinfo {title} {Emerging rare-earth doped
  material platforms for quantum nanophotonics},\ }\href@noop {} {\bibfield
  {journal} {\bibinfo  {journal} {Nanophotonics}\ }\textbf {\bibinfo {volume}
  {8}},\ \bibinfo {pages} {2003} (\bibinfo {year} {2019})}\BibitemShut
  {NoStop}%
\bibitem [{\citenamefont {Liu}\ and\ \citenamefont
  {Jacquier}(2006)}]{REI-Book}%
  \BibitemOpen
  \bibfield  {author} {\bibinfo {author} {\bibfnamefont {G.}~\bibnamefont
  {Liu}}\ and\ \bibinfo {author} {\bibfnamefont {B.}~\bibnamefont {Jacquier}},\
  }\href@noop {} {\emph {\bibinfo {title} {Spectroscopic properties of rare
  earths in optical materials}}},\ Vol.~\bibinfo {volume} {83}\ (\bibinfo
  {publisher} {Springer Science \& Business Media},\ \bibinfo {year}
  {2006})\BibitemShut {NoStop}%
\bibitem [{\citenamefont {Saglamyurek}\ \emph {et~al.}(2016)\citenamefont
  {Saglamyurek}, \citenamefont {Puigibert}, \citenamefont {Zhou}, \citenamefont
  {Giner}, \citenamefont {Marsili}, \citenamefont {Verma}, \citenamefont {Nam},
  \citenamefont {Oesterling}, \citenamefont {Nippa}, \citenamefont {Oblak}
  \emph {et~al.}}]{saglamyurek2016multiplexed}%
  \BibitemOpen
  \bibfield  {author} {\bibinfo {author} {\bibfnamefont {E.}~\bibnamefont
  {Saglamyurek}}, \bibinfo {author} {\bibfnamefont {M.~G.}\ \bibnamefont
  {Puigibert}}, \bibinfo {author} {\bibfnamefont {Q.}~\bibnamefont {Zhou}},
  \bibinfo {author} {\bibfnamefont {L.}~\bibnamefont {Giner}}, \bibinfo
  {author} {\bibfnamefont {F.}~\bibnamefont {Marsili}}, \bibinfo {author}
  {\bibfnamefont {V.~B.}\ \bibnamefont {Verma}}, \bibinfo {author}
  {\bibfnamefont {S.~W.}\ \bibnamefont {Nam}}, \bibinfo {author} {\bibfnamefont
  {L.}~\bibnamefont {Oesterling}}, \bibinfo {author} {\bibfnamefont
  {D.}~\bibnamefont {Nippa}}, \bibinfo {author} {\bibfnamefont
  {D.}~\bibnamefont {Oblak}}, \emph {et~al.},\ }\bibfield  {title} {\bibinfo
  {title} {A multiplexed light-matter interface for fibre-based quantum
  networks},\ }\href@noop {} {\bibfield  {journal} {\bibinfo  {journal} {Nature
  communications}\ }\textbf {\bibinfo {volume} {7}},\ \bibinfo {pages} {11202}
  (\bibinfo {year} {2016})}\BibitemShut {NoStop}%
\bibitem [{\citenamefont {Nunn}\ \emph {et~al.}(2008)\citenamefont {Nunn},
  \citenamefont {Reim}, \citenamefont {Lee}, \citenamefont {Lorenz},
  \citenamefont {Sussman}, \citenamefont {Walmsley},\ and\ \citenamefont
  {Jaksch}}]{nunn2008multimode}%
  \BibitemOpen
  \bibfield  {author} {\bibinfo {author} {\bibfnamefont {J.}~\bibnamefont
  {Nunn}}, \bibinfo {author} {\bibfnamefont {K.}~\bibnamefont {Reim}}, \bibinfo
  {author} {\bibfnamefont {K.}~\bibnamefont {Lee}}, \bibinfo {author}
  {\bibfnamefont {V.}~\bibnamefont {Lorenz}}, \bibinfo {author} {\bibfnamefont
  {B.}~\bibnamefont {Sussman}}, \bibinfo {author} {\bibfnamefont
  {I.}~\bibnamefont {Walmsley}},\ and\ \bibinfo {author} {\bibfnamefont
  {D.}~\bibnamefont {Jaksch}},\ }\bibfield  {title} {\bibinfo {title}
  {Multimode memories in atomic ensembles},\ }\href@noop {} {\bibfield
  {journal} {\bibinfo  {journal} {Physical Review Letters}\ }\textbf {\bibinfo
  {volume} {101}},\ \bibinfo {pages} {260502} (\bibinfo {year}
  {2008})}\BibitemShut {NoStop}%
\bibitem [{\citenamefont {Preston}(1996)}]{preston1996doppler}%
  \BibitemOpen
  \bibfield  {author} {\bibinfo {author} {\bibfnamefont {D.~W.}\ \bibnamefont
  {Preston}},\ }\bibfield  {title} {\bibinfo {title} {Doppler-free saturated
  absorption: Laser spectroscopy},\ }\href@noop {} {\bibfield  {journal}
  {\bibinfo  {journal} {American Journal of Physics}\ }\textbf {\bibinfo
  {volume} {64}},\ \bibinfo {pages} {1432} (\bibinfo {year}
  {1996})}\BibitemShut {NoStop}%
\bibitem [{\citenamefont {Sinclair}\ \emph {et~al.}(2014)\citenamefont
  {Sinclair}, \citenamefont {Saglamyurek}, \citenamefont {Mallahzadeh},
  \citenamefont {Slater}, \citenamefont {George}, \citenamefont {Ricken},
  \citenamefont {Hedges}, \citenamefont {Oblak}, \citenamefont {Simon},
  \citenamefont {Sohler},\ and\ \citenamefont {Tittel}}]{Multiplexing}%
  \BibitemOpen
  \bibfield  {author} {\bibinfo {author} {\bibfnamefont {N.}~\bibnamefont
  {Sinclair}}, \bibinfo {author} {\bibfnamefont {E.}~\bibnamefont
  {Saglamyurek}}, \bibinfo {author} {\bibfnamefont {H.}~\bibnamefont
  {Mallahzadeh}}, \bibinfo {author} {\bibfnamefont {J.~A.}\ \bibnamefont
  {Slater}}, \bibinfo {author} {\bibfnamefont {M.}~\bibnamefont {George}},
  \bibinfo {author} {\bibfnamefont {R.}~\bibnamefont {Ricken}}, \bibinfo
  {author} {\bibfnamefont {M.~P.}\ \bibnamefont {Hedges}}, \bibinfo {author}
  {\bibfnamefont {D.}~\bibnamefont {Oblak}}, \bibinfo {author} {\bibfnamefont
  {C.}~\bibnamefont {Simon}}, \bibinfo {author} {\bibfnamefont
  {W.}~\bibnamefont {Sohler}},\ and\ \bibinfo {author} {\bibfnamefont
  {W.}~\bibnamefont {Tittel}},\ }\bibfield  {title} {\bibinfo {title} {Spectral
  multiplexing for scalable quantum photonics using an atomic frequency comb
  quantum memory and feed-forward control},\ }\href
  {https://doi.org/10.1103/PhysRevLett.113.053603} {\bibfield  {journal}
  {\bibinfo  {journal} {Phys. Rev. Lett.}\ }\textbf {\bibinfo {volume} {113}},\
  \bibinfo {pages} {053603} (\bibinfo {year} {2014})}\BibitemShut {NoStop}%
\bibitem [{\citenamefont {Puigibert}\ \emph {et~al.}(2020)\citenamefont
  {Puigibert}, \citenamefont {Askarani}, \citenamefont {Davidson},
  \citenamefont {Verma}, \citenamefont {Shaw}, \citenamefont {Nam},
  \citenamefont {Lutz}, \citenamefont {Amaral}, \citenamefont {Oblak},\ and\
  \citenamefont {Tittel}}]{Entanglement_and_nonlocality}%
  \BibitemOpen
  \bibfield  {author} {\bibinfo {author} {\bibfnamefont {M.~l.~G.}\
  \bibnamefont {Puigibert}}, \bibinfo {author} {\bibfnamefont {M.~F.}\
  \bibnamefont {Askarani}}, \bibinfo {author} {\bibfnamefont {J.~H.}\
  \bibnamefont {Davidson}}, \bibinfo {author} {\bibfnamefont {V.~B.}\
  \bibnamefont {Verma}}, \bibinfo {author} {\bibfnamefont {M.~D.}\ \bibnamefont
  {Shaw}}, \bibinfo {author} {\bibfnamefont {S.~W.}\ \bibnamefont {Nam}},
  \bibinfo {author} {\bibfnamefont {T.}~\bibnamefont {Lutz}}, \bibinfo {author}
  {\bibfnamefont {G.~C.}\ \bibnamefont {Amaral}}, \bibinfo {author}
  {\bibfnamefont {D.}~\bibnamefont {Oblak}},\ and\ \bibinfo {author}
  {\bibfnamefont {W.}~\bibnamefont {Tittel}},\ }\bibfield  {title} {\bibinfo
  {title} {Entanglement and nonlocality between disparate solid-state quantum
  memories mediated by photons},\ }\href
  {https://doi.org/10.1103/PhysRevResearch.2.013039} {\bibfield  {journal}
  {\bibinfo  {journal} {Phys. Rev. Research}\ }\textbf {\bibinfo {volume}
  {2}},\ \bibinfo {pages} {013039} (\bibinfo {year} {2020})}\BibitemShut
  {NoStop}%
\bibitem [{\citenamefont {B{\"o}ttger}\ \emph {et~al.}(2006)\citenamefont
  {B{\"o}ttger}, \citenamefont {Thiel}, \citenamefont {Sun},\ and\
  \citenamefont {Cone}}]{bottger2006optical}%
  \BibitemOpen
  \bibfield  {author} {\bibinfo {author} {\bibfnamefont {T.}~\bibnamefont
  {B{\"o}ttger}}, \bibinfo {author} {\bibfnamefont {C.}~\bibnamefont {Thiel}},
  \bibinfo {author} {\bibfnamefont {Y.}~\bibnamefont {Sun}},\ and\ \bibinfo
  {author} {\bibfnamefont {R.}~\bibnamefont {Cone}},\ }\bibfield  {title}
  {\bibinfo {title} {Optical decoherence and spectral diffusion at 1.5 $\mu$ m
  in er 3+: Y 2 sio 5 versus magnetic field, temperature, and er 3+
  concentration},\ }\href@noop {} {\bibfield  {journal} {\bibinfo  {journal}
  {Physical Review B}\ }\textbf {\bibinfo {volume} {73}},\ \bibinfo {pages}
  {075101} (\bibinfo {year} {2006})}\BibitemShut {NoStop}%
\bibitem [{\citenamefont {Das}\ \emph {et~al.}(2023)\citenamefont {Das},
  \citenamefont {Davidson}, \citenamefont {Chakraborty}, \citenamefont
  {Tchebotareva},\ and\ \citenamefont {Tittel}}]{Das:23}%
  \BibitemOpen
  \bibfield  {author} {\bibinfo {author} {\bibfnamefont {A.}~\bibnamefont
  {Das}}, \bibinfo {author} {\bibfnamefont {J.~H.}\ \bibnamefont {Davidson}},
  \bibinfo {author} {\bibfnamefont {T.}~\bibnamefont {Chakraborty}}, \bibinfo
  {author} {\bibfnamefont {A.~L.}\ \bibnamefont {Tchebotareva}},\ and\ \bibinfo
  {author} {\bibfnamefont {W.}~\bibnamefont {Tittel}},\ }\bibfield  {title}
  {\bibinfo {title} {Towards an alignment-free, impedance-matched cavity
  quantum memory in a thulium-doped crystal},\ }in\ \href
  {https://doi.org/10.1364/CLEO_SI.2023.STh5C.6} {\emph {\bibinfo {booktitle}
  {CLEO 2023}}}\ (\bibinfo  {publisher} {Optica Publishing Group},\ \bibinfo
  {year} {2023})\ p.\ \bibinfo {pages} {STh5C.6}\BibitemShut {NoStop}%
\bibitem [{\citenamefont {Sinclair}\ \emph {et~al.}(2017)\citenamefont
  {Sinclair}, \citenamefont {Oblak}, \citenamefont {Thiel}, \citenamefont
  {Cone},\ and\ \citenamefont {Tittel}}]{sinclair2017properties}%
  \BibitemOpen
  \bibfield  {author} {\bibinfo {author} {\bibfnamefont {N.}~\bibnamefont
  {Sinclair}}, \bibinfo {author} {\bibfnamefont {D.}~\bibnamefont {Oblak}},
  \bibinfo {author} {\bibfnamefont {C.~W.}\ \bibnamefont {Thiel}}, \bibinfo
  {author} {\bibfnamefont {R.~L.}\ \bibnamefont {Cone}},\ and\ \bibinfo
  {author} {\bibfnamefont {W.}~\bibnamefont {Tittel}},\ }\bibfield  {title}
  {\bibinfo {title} {Properties of a rare-earth-ion-doped waveguide at
  sub-kelvin temperatures for quantum signal processing},\ }\href@noop {}
  {\bibfield  {journal} {\bibinfo  {journal} {Physical review letters}\
  }\textbf {\bibinfo {volume} {118}},\ \bibinfo {pages} {100504} (\bibinfo
  {year} {2017})}\BibitemShut {NoStop}%
\bibitem [{\citenamefont {Sun}\ \emph {et~al.}(2012)\citenamefont {Sun},
  \citenamefont {Thiel},\ and\ \citenamefont {Cone}}]{sun2012optical}%
  \BibitemOpen
  \bibfield  {author} {\bibinfo {author} {\bibfnamefont {Y.}~\bibnamefont
  {Sun}}, \bibinfo {author} {\bibfnamefont {C.}~\bibnamefont {Thiel}},\ and\
  \bibinfo {author} {\bibfnamefont {R.}~\bibnamefont {Cone}},\ }\bibfield
  {title} {\bibinfo {title} {Optical decoherence and energy level structure of
  0.1\% tm 3+: Linbo 3},\ }\href@noop {} {\bibfield  {journal} {\bibinfo
  {journal} {Physical Review B}\ }\textbf {\bibinfo {volume} {85}},\ \bibinfo
  {pages} {165106} (\bibinfo {year} {2012})}\BibitemShut {NoStop}%
\bibitem [{\citenamefont {Sinclair}\ \emph {et~al.}(2021)\citenamefont
  {Sinclair}, \citenamefont {Oblak}, \citenamefont {Saglamyurek}, \citenamefont
  {Cone}, \citenamefont {Thiel},\ and\ \citenamefont
  {Tittel}}]{sinclair2021optical}%
  \BibitemOpen
  \bibfield  {author} {\bibinfo {author} {\bibfnamefont {N.}~\bibnamefont
  {Sinclair}}, \bibinfo {author} {\bibfnamefont {D.}~\bibnamefont {Oblak}},
  \bibinfo {author} {\bibfnamefont {E.}~\bibnamefont {Saglamyurek}}, \bibinfo
  {author} {\bibfnamefont {R.~L.}\ \bibnamefont {Cone}}, \bibinfo {author}
  {\bibfnamefont {C.~W.}\ \bibnamefont {Thiel}},\ and\ \bibinfo {author}
  {\bibfnamefont {W.}~\bibnamefont {Tittel}},\ }\bibfield  {title} {\bibinfo
  {title} {Optical coherence and energy-level properties of a tm 3+-doped li nb
  o 3 waveguide at subkelvin temperatures},\ }\href@noop {} {\bibfield
  {journal} {\bibinfo  {journal} {Physical Review B}\ }\textbf {\bibinfo
  {volume} {103}},\ \bibinfo {pages} {134105} (\bibinfo {year}
  {2021})}\BibitemShut {NoStop}%
\bibitem [{\citenamefont {Saglamyurek}\ \emph {et~al.}(2012)\citenamefont
  {Saglamyurek}, \citenamefont {Sinclair}, \citenamefont {Jin}, \citenamefont
  {Slater}, \citenamefont {Oblak}, \citenamefont {Bussieres}, \citenamefont
  {George}, \citenamefont {Ricken}, \citenamefont {Sohler},\ and\ \citenamefont
  {Tittel}}]{saglamyurek2012conditional}%
  \BibitemOpen
  \bibfield  {author} {\bibinfo {author} {\bibfnamefont {E.}~\bibnamefont
  {Saglamyurek}}, \bibinfo {author} {\bibfnamefont {N.}~\bibnamefont
  {Sinclair}}, \bibinfo {author} {\bibfnamefont {J.}~\bibnamefont {Jin}},
  \bibinfo {author} {\bibfnamefont {J.~A.}\ \bibnamefont {Slater}}, \bibinfo
  {author} {\bibfnamefont {D.}~\bibnamefont {Oblak}}, \bibinfo {author}
  {\bibfnamefont {F.}~\bibnamefont {Bussieres}}, \bibinfo {author}
  {\bibfnamefont {M.}~\bibnamefont {George}}, \bibinfo {author} {\bibfnamefont
  {R.}~\bibnamefont {Ricken}}, \bibinfo {author} {\bibfnamefont
  {W.}~\bibnamefont {Sohler}},\ and\ \bibinfo {author} {\bibfnamefont
  {W.}~\bibnamefont {Tittel}},\ }\bibfield  {title} {\bibinfo {title}
  {Conditional detection of pure quantum states of light after storage in a
  tm-doped waveguide},\ }\href@noop {} {\bibfield  {journal} {\bibinfo
  {journal} {Physical Review Letters}\ }\textbf {\bibinfo {volume} {108}},\
  \bibinfo {pages} {083602} (\bibinfo {year} {2012})}\BibitemShut {NoStop}%
\bibitem [{\citenamefont {McAuslan}\ \emph {et~al.}(2012)\citenamefont
  {McAuslan}, \citenamefont {Bartholomew}, \citenamefont {Sellars},\ and\
  \citenamefont {Longdell}}]{mcauslan2012reducing}%
  \BibitemOpen
  \bibfield  {author} {\bibinfo {author} {\bibfnamefont {D.}~\bibnamefont
  {McAuslan}}, \bibinfo {author} {\bibfnamefont {J.}~\bibnamefont
  {Bartholomew}}, \bibinfo {author} {\bibfnamefont {M.}~\bibnamefont
  {Sellars}},\ and\ \bibinfo {author} {\bibfnamefont {J.~J.}\ \bibnamefont
  {Longdell}},\ }\bibfield  {title} {\bibinfo {title} {Reducing decoherence in
  optical and spin transitions in rare-earth-metal-ion--doped materials},\
  }\href@noop {} {\bibfield  {journal} {\bibinfo  {journal} {Physical Review
  A}\ }\textbf {\bibinfo {volume} {85}},\ \bibinfo {pages} {032339} (\bibinfo
  {year} {2012})}\BibitemShut {NoStop}%
\bibitem [{\citenamefont {Davidson}\ \emph {et~al.}(2021)\citenamefont
  {Davidson}, \citenamefont {Woodburn}, \citenamefont {Marsh}, \citenamefont
  {Olson}, \citenamefont {Olivera}, \citenamefont {Das}, \citenamefont
  {Askarani}, \citenamefont {Tittel}, \citenamefont {Cone},\ and\ \citenamefont
  {Thiel}}]{Davidson}%
  \BibitemOpen
  \bibfield  {author} {\bibinfo {author} {\bibfnamefont {J.~H.}\ \bibnamefont
  {Davidson}}, \bibinfo {author} {\bibfnamefont {P.~J.~T.}\ \bibnamefont
  {Woodburn}}, \bibinfo {author} {\bibfnamefont {A.~D.}\ \bibnamefont {Marsh}},
  \bibinfo {author} {\bibfnamefont {K.~J.}\ \bibnamefont {Olson}}, \bibinfo
  {author} {\bibfnamefont {A.}~\bibnamefont {Olivera}}, \bibinfo {author}
  {\bibfnamefont {A.}~\bibnamefont {Das}}, \bibinfo {author} {\bibfnamefont
  {M.~F.}\ \bibnamefont {Askarani}}, \bibinfo {author} {\bibfnamefont
  {W.}~\bibnamefont {Tittel}}, \bibinfo {author} {\bibfnamefont {R.~L.}\
  \bibnamefont {Cone}},\ and\ \bibinfo {author} {\bibfnamefont {C.~W.}\
  \bibnamefont {Thiel}},\ }\bibfield  {title} {\bibinfo {title} {Measurement of
  the thulium ion spin hamiltonian in an yttrium gallium garnet host crystal},\
  }\href {https://doi.org/10.1103/PhysRevB.104.134103} {\bibfield  {journal}
  {\bibinfo  {journal} {Phys. Rev. B}\ }\textbf {\bibinfo {volume} {104}},\
  \bibinfo {pages} {134103} (\bibinfo {year} {2021})}\BibitemShut {NoStop}%
\bibitem [{\citenamefont {Thiel}\ \emph {et~al.}(2012)\citenamefont {Thiel},
  \citenamefont {Babbitt},\ and\ \citenamefont {Cone}}]{Thiel_strain}%
  \BibitemOpen
  \bibfield  {author} {\bibinfo {author} {\bibfnamefont {C.~W.}\ \bibnamefont
  {Thiel}}, \bibinfo {author} {\bibfnamefont {W.~R.}\ \bibnamefont {Babbitt}},\
  and\ \bibinfo {author} {\bibfnamefont {R.~L.}\ \bibnamefont {Cone}},\
  }\bibfield  {title} {\bibinfo {title} {Optical decoherence studies of yttrium
  oxyorthosilicate y${}_{2}$sio${}_{5}$ codoped with er${}^{3+}$ and
  eu${}^{3+}$ for optical signal processing and quantum information
  applications at 1.5 microns},\ }\href
  {https://doi.org/10.1103/PhysRevB.85.174302} {\bibfield  {journal} {\bibinfo
  {journal} {Phys. Rev. B}\ }\textbf {\bibinfo {volume} {85}},\ \bibinfo
  {pages} {174302} (\bibinfo {year} {2012})}\BibitemShut {NoStop}%
\bibitem [{\citenamefont {B\"ottger}\ \emph {et~al.}(2008)\citenamefont
  {B\"ottger}, \citenamefont {Thiel}, \citenamefont {Cone},\ and\ \citenamefont
  {Sun}}]{Bottger_codoping}%
  \BibitemOpen
  \bibfield  {author} {\bibinfo {author} {\bibfnamefont {T.}~\bibnamefont
  {B\"ottger}}, \bibinfo {author} {\bibfnamefont {C.~W.}\ \bibnamefont
  {Thiel}}, \bibinfo {author} {\bibfnamefont {R.~L.}\ \bibnamefont {Cone}},\
  and\ \bibinfo {author} {\bibfnamefont {Y.}~\bibnamefont {Sun}},\ }\bibfield
  {title} {\bibinfo {title} {Controlled compositional disorder in
  ${\mathrm{er}}^{3+}:{\mathrm{y}}_{2}\mathrm{Si}{\mathrm{o}}_{5}$ provides a
  wide-bandwidth spectral hole burning material at
  $1.5\phantom{\rule{0.3em}{0ex}}\ensuremath{\mu}\mathrm{m}$},\ }\href
  {https://doi.org/10.1103/PhysRevB.77.155125} {\bibfield  {journal} {\bibinfo
  {journal} {Phys. Rev. B}\ }\textbf {\bibinfo {volume} {77}},\ \bibinfo
  {pages} {155125} (\bibinfo {year} {2008})}\BibitemShut {NoStop}%
\bibitem [{\citenamefont {Thiel}\ \emph {et~al.}(2014)\citenamefont {Thiel},
  \citenamefont {Sinclair}, \citenamefont {Tittel},\ and\ \citenamefont
  {Cone}}]{thiel2014optical}%
  \BibitemOpen
  \bibfield  {author} {\bibinfo {author} {\bibfnamefont {C.~W.}\ \bibnamefont
  {Thiel}}, \bibinfo {author} {\bibfnamefont {N.}~\bibnamefont {Sinclair}},
  \bibinfo {author} {\bibfnamefont {W.}~\bibnamefont {Tittel}},\ and\ \bibinfo
  {author} {\bibfnamefont {R.~L.}\ \bibnamefont {Cone}},\ }\bibfield  {title}
  {\bibinfo {title} {Optical decoherence studies of tm 3+: Y 3 ga 5 o 12},\
  }\href@noop {} {\bibfield  {journal} {\bibinfo  {journal} {Physical Review
  B}\ }\textbf {\bibinfo {volume} {90}},\ \bibinfo {pages} {214301} (\bibinfo
  {year} {2014})}\BibitemShut {NoStop}%
\bibitem [{\citenamefont {Ferrier}\ \emph {et~al.}(2018)\citenamefont
  {Ferrier}, \citenamefont {Ilas}, \citenamefont {Goldner},\ and\ \citenamefont
  {Louchet-Chauvet}}]{ferrier2018scandium}%
  \BibitemOpen
  \bibfield  {author} {\bibinfo {author} {\bibfnamefont {A.}~\bibnamefont
  {Ferrier}}, \bibinfo {author} {\bibfnamefont {S.}~\bibnamefont {Ilas}},
  \bibinfo {author} {\bibfnamefont {P.}~\bibnamefont {Goldner}},\ and\ \bibinfo
  {author} {\bibfnamefont {A.}~\bibnamefont {Louchet-Chauvet}},\ }\bibfield
  {title} {\bibinfo {title} {Scandium doped tm: Yag ceramics and single
  crystals: Coherent and high resolution spectroscopy},\ }\href@noop {}
  {\bibfield  {journal} {\bibinfo  {journal} {Journal of Luminescence}\
  }\textbf {\bibinfo {volume} {194}},\ \bibinfo {pages} {116} (\bibinfo {year}
  {2018})}\BibitemShut {NoStop}%
\bibitem [{\citenamefont {Zhang}\ \emph {et~al.}(2020)\citenamefont {Zhang},
  \citenamefont {Louchet-Chauvet}, \citenamefont {Morvan}, \citenamefont
  {Berger}, \citenamefont {Goldner},\ and\ \citenamefont
  {Ferrier}}]{zhang2020tailoring}%
  \BibitemOpen
  \bibfield  {author} {\bibinfo {author} {\bibfnamefont {Z.}~\bibnamefont
  {Zhang}}, \bibinfo {author} {\bibfnamefont {A.}~\bibnamefont
  {Louchet-Chauvet}}, \bibinfo {author} {\bibfnamefont {L.}~\bibnamefont
  {Morvan}}, \bibinfo {author} {\bibfnamefont {P.}~\bibnamefont {Berger}},
  \bibinfo {author} {\bibfnamefont {P.}~\bibnamefont {Goldner}},\ and\ \bibinfo
  {author} {\bibfnamefont {A.}~\bibnamefont {Ferrier}},\ }\bibfield  {title}
  {\bibinfo {title} {Tailoring the 3f4 level lifetime in tm3+: Y3al5o12 by eu3+
  co-doping for signal processing application},\ }\href@noop {} {\bibfield
  {journal} {\bibinfo  {journal} {Journal of Luminescence}\ }\textbf {\bibinfo
  {volume} {222}},\ \bibinfo {pages} {117107} (\bibinfo {year}
  {2020})}\BibitemShut {NoStop}%
\bibitem [{\citenamefont {Bennett}\ and\ \citenamefont
  {Brassard}(2014)}]{Bennett2014}%
  \BibitemOpen
  \bibfield  {author} {\bibinfo {author} {\bibfnamefont {C.~H.}\ \bibnamefont
  {Bennett}}\ and\ \bibinfo {author} {\bibfnamefont {G.}~\bibnamefont
  {Brassard}},\ }\bibfield  {title} {\bibinfo {title} {Quantum cryptography:
  {{Public}} key distribution and coin tossing},\ }\href
  {https://doi.org/10.1016/j.tcs.2014.05.025} {\bibfield  {journal} {\bibinfo
  {journal} {Theoretical Computer Science}\ }\textbf {\bibinfo {volume}
  {560}},\ \bibinfo {pages} {7} (\bibinfo {year} {2014})}\BibitemShut {NoStop}%
\bibitem [{\citenamefont {Lo}\ \emph {et~al.}(2005)\citenamefont {Lo},
  \citenamefont {Chau},\ and\ \citenamefont {Ardehali}}]{Lo2005}%
  \BibitemOpen
  \bibfield  {author} {\bibinfo {author} {\bibfnamefont {H.-K.}\ \bibnamefont
  {Lo}}, \bibinfo {author} {\bibfnamefont {H.~F.}\ \bibnamefont {Chau}},\ and\
  \bibinfo {author} {\bibfnamefont {M.}~\bibnamefont {Ardehali}},\ }\bibfield
  {title} {\bibinfo {title} {Efficient quantum key distribution scheme and a
  proof of its unconditional security},\ }\href
  {https://doi.org/10.1007/s00145-004-0142-y} {\bibfield  {journal} {\bibinfo
  {journal} {Journal of Cryptology}\ }\textbf {\bibinfo {volume} {18}},\
  \bibinfo {pages} {133} (\bibinfo {year} {2005})}\BibitemShut {NoStop}%
\bibitem [{\citenamefont {Scarani}\ \emph {et~al.}(2009)\citenamefont
  {Scarani}, \citenamefont {Bechmann-Pasquinucci}, \citenamefont {Cerf},
  \citenamefont {Du\ifmmode~\check{s}\else \v{s}\fi{}ek}, \citenamefont
  {L\"utkenhaus},\ and\ \citenamefont {Peev}}]{Scarani2009}%
  \BibitemOpen
  \bibfield  {author} {\bibinfo {author} {\bibfnamefont {V.}~\bibnamefont
  {Scarani}}, \bibinfo {author} {\bibfnamefont {H.}~\bibnamefont
  {Bechmann-Pasquinucci}}, \bibinfo {author} {\bibfnamefont {N.~J.}\
  \bibnamefont {Cerf}}, \bibinfo {author} {\bibfnamefont {M.}~\bibnamefont
  {Du\ifmmode~\check{s}\else \v{s}\fi{}ek}}, \bibinfo {author} {\bibfnamefont
  {N.}~\bibnamefont {L\"utkenhaus}},\ and\ \bibinfo {author} {\bibfnamefont
  {M.}~\bibnamefont {Peev}},\ }\bibfield  {title} {\bibinfo {title} {The
  security of practical quantum key distribution},\ }\href
  {https://doi.org/10.1103/RevModPhys.81.1301} {\bibfield  {journal} {\bibinfo
  {journal} {Rev. Mod. Phys.}\ }\textbf {\bibinfo {volume} {81}},\ \bibinfo
  {pages} {1301} (\bibinfo {year} {2009})}\BibitemShut {NoStop}%
\bibitem [{\citenamefont {Huber}\ \emph {et~al.}(2018)\citenamefont {Huber},
  \citenamefont {Reindl}, \citenamefont {Aberl}, \citenamefont {Rastelli},\
  and\ \citenamefont {Trotta}}]{Huber2018}%
  \BibitemOpen
  \bibfield  {author} {\bibinfo {author} {\bibfnamefont {D.}~\bibnamefont
  {Huber}}, \bibinfo {author} {\bibfnamefont {M.}~\bibnamefont {Reindl}},
  \bibinfo {author} {\bibfnamefont {J.}~\bibnamefont {Aberl}}, \bibinfo
  {author} {\bibfnamefont {A.}~\bibnamefont {Rastelli}},\ and\ \bibinfo
  {author} {\bibfnamefont {R.}~\bibnamefont {Trotta}},\ }\bibfield  {title}
  {\bibinfo {title} {Semiconductor quantum dots as an ideal source of
  polarization-entangled photon pairs on-demand: a review},\ }\href
  {https://doi.org/10.1088/2040-8986/aac4c4} {\bibfield  {journal} {\bibinfo
  {journal} {Journal of Optics}\ }\textbf {\bibinfo {volume} {20}},\ \bibinfo
  {pages} {073002} (\bibinfo {year} {2018})}\BibitemShut {NoStop}%
\bibitem [{\citenamefont {Heller}\ \emph {et~al.}(2022)\citenamefont {Heller},
  \citenamefont {Lowinski}, \citenamefont {Theophilo}, \citenamefont
  {Padr\'on-Brito},\ and\ \citenamefont
  {de~Riedmatten}}]{PhysRevApplied.18.024036}%
  \BibitemOpen
  \bibfield  {author} {\bibinfo {author} {\bibfnamefont {L.}~\bibnamefont
  {Heller}}, \bibinfo {author} {\bibfnamefont {J.}~\bibnamefont {Lowinski}},
  \bibinfo {author} {\bibfnamefont {K.}~\bibnamefont {Theophilo}}, \bibinfo
  {author} {\bibfnamefont {A.}~\bibnamefont {Padr\'on-Brito}},\ and\ \bibinfo
  {author} {\bibfnamefont {H.}~\bibnamefont {de~Riedmatten}},\ }\bibfield
  {title} {\bibinfo {title} {Raman storage of quasideterministic single photons
  generated by rydberg collective excitations in a low-noise quantum memory},\
  }\href {https://doi.org/10.1103/PhysRevApplied.18.024036} {\bibfield
  {journal} {\bibinfo  {journal} {Phys. Rev. Appl.}\ }\textbf {\bibinfo
  {volume} {18}},\ \bibinfo {pages} {024036} (\bibinfo {year}
  {2022})}\BibitemShut {NoStop}%
\bibitem [{\citenamefont {Guo}\ \emph {et~al.}(2023)\citenamefont {Guo},
  \citenamefont {Liu}, \citenamefont {Sun}, \citenamefont {Ren}, \citenamefont
  {Wang},\ and\ \citenamefont {Zhong}}]{guo2023rare}%
  \BibitemOpen
  \bibfield  {author} {\bibinfo {author} {\bibfnamefont {M.}~\bibnamefont
  {Guo}}, \bibinfo {author} {\bibfnamefont {S.}~\bibnamefont {Liu}}, \bibinfo
  {author} {\bibfnamefont {W.}~\bibnamefont {Sun}}, \bibinfo {author}
  {\bibfnamefont {M.}~\bibnamefont {Ren}}, \bibinfo {author} {\bibfnamefont
  {F.}~\bibnamefont {Wang}},\ and\ \bibinfo {author} {\bibfnamefont
  {M.}~\bibnamefont {Zhong}},\ }\bibfield  {title} {\bibinfo {title}
  {Rare-earth quantum memories: The experimental status quo},\ }\href@noop {}
  {\bibfield  {journal} {\bibinfo  {journal} {Frontiers of Physics}\ }\textbf
  {\bibinfo {volume} {18}},\ \bibinfo {pages} {21303} (\bibinfo {year}
  {2023})}\BibitemShut {NoStop}%
\bibitem [{\citenamefont {Tchebotareva}\ \emph {et~al.}(2019)\citenamefont
  {Tchebotareva}, \citenamefont {Hermans}, \citenamefont {Humphreys},
  \citenamefont {Voigt}, \citenamefont {Harmsma}, \citenamefont {Cheng},
  \citenamefont {Verlaan}, \citenamefont {Dijkhuizen}, \citenamefont {de~Jong},
  \citenamefont {Dr\'eau},\ and\ \citenamefont {Hanson}}]{Tchebotareva2019}%
  \BibitemOpen
  \bibfield  {author} {\bibinfo {author} {\bibfnamefont {A.}~\bibnamefont
  {Tchebotareva}}, \bibinfo {author} {\bibfnamefont {S.~L.~N.}\ \bibnamefont
  {Hermans}}, \bibinfo {author} {\bibfnamefont {P.~C.}\ \bibnamefont
  {Humphreys}}, \bibinfo {author} {\bibfnamefont {D.}~\bibnamefont {Voigt}},
  \bibinfo {author} {\bibfnamefont {P.~J.}\ \bibnamefont {Harmsma}}, \bibinfo
  {author} {\bibfnamefont {L.~K.}\ \bibnamefont {Cheng}}, \bibinfo {author}
  {\bibfnamefont {A.~L.}\ \bibnamefont {Verlaan}}, \bibinfo {author}
  {\bibfnamefont {N.}~\bibnamefont {Dijkhuizen}}, \bibinfo {author}
  {\bibfnamefont {W.}~\bibnamefont {de~Jong}}, \bibinfo {author} {\bibfnamefont
  {A.}~\bibnamefont {Dr\'eau}},\ and\ \bibinfo {author} {\bibfnamefont
  {R.}~\bibnamefont {Hanson}},\ }\bibfield  {title} {\bibinfo {title}
  {Entanglement between a diamond spin qubit and a photonic time-bin qubit at
  telecom wavelength},\ }\href {https://doi.org/10.1103/PhysRevLett.123.063601}
  {\bibfield  {journal} {\bibinfo  {journal} {Phys. Rev. Lett.}\ }\textbf
  {\bibinfo {volume} {123}},\ \bibinfo {pages} {063601} (\bibinfo {year}
  {2019})}\BibitemShut {NoStop}%
\bibitem [{\citenamefont {Iuliano}\ \emph {et~al.}(2023)\citenamefont
  {Iuliano}, \citenamefont {Roehsner}, \citenamefont {Alifasi}, \citenamefont
  {Chakraborty}, \citenamefont {Stolk}, \citenamefont {Weaver}, \citenamefont
  {Sholkina}, \citenamefont {Loukiantchenko}, \citenamefont {do~Amaral},
  \citenamefont {Tittel},\ and\ \citenamefont {Hanson}}]{Iuliano2023}%
  \BibitemOpen
  \bibfield  {author} {\bibinfo {author} {\bibfnamefont {M.}~\bibnamefont
  {Iuliano}}, \bibinfo {author} {\bibfnamefont {M.-C.}\ \bibnamefont
  {Roehsner}}, \bibinfo {author} {\bibfnamefont {N.}~\bibnamefont {Alifasi}},
  \bibinfo {author} {\bibfnamefont {T.}~\bibnamefont {Chakraborty}}, \bibinfo
  {author} {\bibfnamefont {A.~J.}\ \bibnamefont {Stolk}}, \bibinfo {author}
  {\bibfnamefont {M.~J.}\ \bibnamefont {Weaver}}, \bibinfo {author}
  {\bibfnamefont {M.~O.}\ \bibnamefont {Sholkina}}, \bibinfo {author}
  {\bibfnamefont {E.}~\bibnamefont {Loukiantchenko}}, \bibinfo {author}
  {\bibfnamefont {G.~C.}\ \bibnamefont {do~Amaral}}, \bibinfo {author}
  {\bibfnamefont {W.}~\bibnamefont {Tittel}},\ and\ \bibinfo {author}
  {\bibfnamefont {R.}~\bibnamefont {Hanson}},\ }\bibfield  {title} {\bibinfo
  {title} {Interfacing an nv-center in diamond and a rare-earth ion compatible
  photonic time-bin qubit},\ }in\ \href
  {https://doi.org/10.1364/QUANTUM.2023.QW4A.7} {\emph {\bibinfo {booktitle}
  {Optica Quantum 2.0 Conference and Exhibition}}}\ (\bibinfo  {publisher}
  {Optica Publishing Group},\ \bibinfo {year} {2023})\ p.\ \bibinfo {pages}
  {QW4A.7}\BibitemShut {NoStop}%
\bibitem [{\citenamefont {Quan}\ and\ \citenamefont {Loncar}(2003)}]{Quan2011}%
  \BibitemOpen
  \bibfield  {author} {\bibinfo {author} {\bibfnamefont {Q.}~\bibnamefont
  {Quan}}\ and\ \bibinfo {author} {\bibfnamefont {M.}~\bibnamefont {Loncar}},\
  }\bibfield  {title} {\bibinfo {title} {Deterministic design of wavelength
  scale, ultra-high q photonic crystal nanobeam cavities},\ }\href@noop {}
  {\bibfield  {journal} {\bibinfo  {journal} {J. Opt. Soc. Am. A}\ }\textbf
  {\bibinfo {volume} {424}},\ \bibinfo {pages} {435} (\bibinfo {year}
  {2003})}\BibitemShut {NoStop}%
\bibitem [{\citenamefont {Goban}\ \emph {et~al.}(2014)\citenamefont {Goban},
  \citenamefont {Hung}, \citenamefont {Yu}, \citenamefont {Hood}, \citenamefont
  {Muniz}, \citenamefont {Lee}, \citenamefont {Martin}, \citenamefont
  {McClung}, \citenamefont {Choi}, \citenamefont {Chang}, \citenamefont
  {Painter},\ and\ \citenamefont {Kimble}}]{Goban2014}%
  \BibitemOpen
  \bibfield  {author} {\bibinfo {author} {\bibfnamefont {A.}~\bibnamefont
  {Goban}}, \bibinfo {author} {\bibfnamefont {C.-L.}\ \bibnamefont {Hung}},
  \bibinfo {author} {\bibfnamefont {S.-P.}\ \bibnamefont {Yu}}, \bibinfo
  {author} {\bibfnamefont {J.}~\bibnamefont {Hood}}, \bibinfo {author}
  {\bibfnamefont {J.}~\bibnamefont {Muniz}}, \bibinfo {author} {\bibfnamefont
  {J.}~\bibnamefont {Lee}}, \bibinfo {author} {\bibfnamefont {M.}~\bibnamefont
  {Martin}}, \bibinfo {author} {\bibfnamefont {A.}~\bibnamefont {McClung}},
  \bibinfo {author} {\bibfnamefont {K.}~\bibnamefont {Choi}}, \bibinfo {author}
  {\bibfnamefont {D.}~\bibnamefont {Chang}}, \bibinfo {author} {\bibfnamefont
  {O.}~\bibnamefont {Painter}},\ and\ \bibinfo {author} {\bibfnamefont
  {H.}~\bibnamefont {Kimble}},\ }\bibfield  {title} {\bibinfo {title}
  {Atom–light interactions in photonic crystals},\ }\href
  {https://doi.org/10.1038/ncomms4808} {\bibfield  {journal} {\bibinfo
  {journal} {Nature Communications 2014 5:1}\ }\textbf {\bibinfo {volume}
  {5}},\ \bibinfo {pages} {1} (\bibinfo {year} {2014})}\BibitemShut {NoStop}%
\bibitem [{\citenamefont {Philipp}(1973)}]{Philipp1973}%
  \BibitemOpen
  \bibfield  {author} {\bibinfo {author} {\bibfnamefont {H.~R.}\ \bibnamefont
  {Philipp}},\ }\bibfield  {title} {\bibinfo {title} {Optical properties of
  silicon nitride},\ }\href {https://doi.org/10.1149/1.2403440} {\bibfield
  {journal} {\bibinfo  {journal} {Journal of The Electrochemical Society}\
  }\textbf {\bibinfo {volume} {120}},\ \bibinfo {pages} {295} (\bibinfo {year}
  {1973})}\BibitemShut {NoStop}%
\bibitem [{\citenamefont {Menon}\ \emph {et~al.}(2020)\citenamefont {Menon},
  \citenamefont {Singh}, \citenamefont {Borregaard},\ and\ \citenamefont
  {Bernien}}]{Menon2020}%
  \BibitemOpen
  \bibfield  {author} {\bibinfo {author} {\bibfnamefont {S.~G.}\ \bibnamefont
  {Menon}}, \bibinfo {author} {\bibfnamefont {K.}~\bibnamefont {Singh}},
  \bibinfo {author} {\bibfnamefont {J.}~\bibnamefont {Borregaard}},\ and\
  \bibinfo {author} {\bibfnamefont {H.}~\bibnamefont {Bernien}},\ }\bibfield
  {title} {\bibinfo {title} {Nanophotonic quantum network node with neutral
  atoms and an integrated telecom interface},\ }\href@noop {} {\bibfield
  {journal} {\bibinfo  {journal} {New Journal of Physics}\ }\textbf {\bibinfo
  {volume} {22}},\ \bibinfo {pages} {073033} (\bibinfo {year}
  {2020})}\BibitemShut {NoStop}%
\end{thebibliography}%

\clearpage
\onecolumngrid
\appendix
\renewcommand\thefigure{\thesection\arabic{figure}}

\section{Rb emitter Hamiltonian} \label{App_Rb_Hamiltonian}
\setcounter{figure}{0}
\setcounter{equation}{0}

\begin{figure}
	\centering
	\captionsetup{justification=centering}
	\includegraphics[width = 0.8\textwidth]{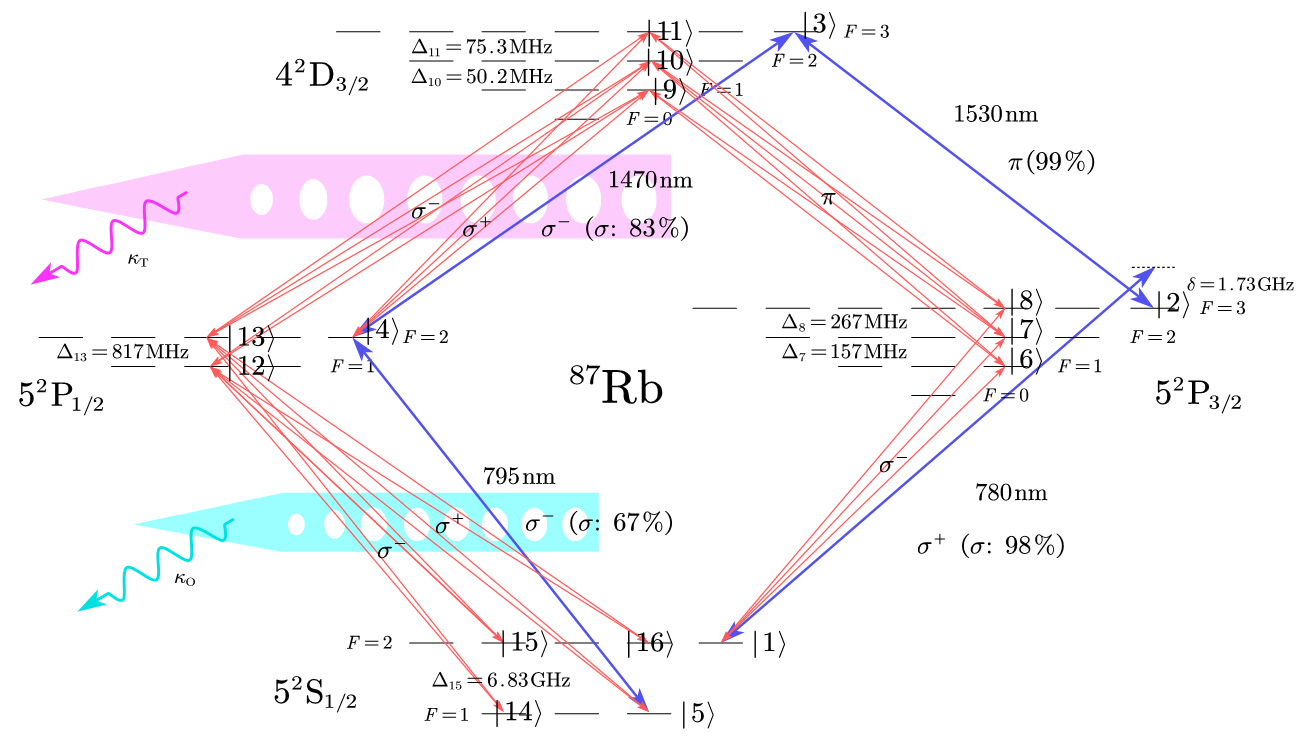}
	\caption{Atomic-level diagram illustrating all the relevant Rb atomic energy levels participating in the light emission procedure. States $|1\rangle$ to $|5\rangle$ are the intended used levels. They are connected with thick blue arrows as their couplings. The first two couplings are laser-induced and the later two are cavity-induced. The thin red arrows indicate the unintended drivings and coupling caused by the used laser and cavities. The choices of polarization ($\pi$, $\sigma^{+}$, $\sigma^{-}$) and wavelengths of light are noted alongside each coupling arrow. There is a detuning $\delta=1.73$ GHz respecting the $|1\rangle-|2\rangle$ energy difference in the first laser driving. The thin red arrows represent the notable faulty couplings induced by the same lasers and cavity modes. These faulty couplings involve other states denoted as $|6\rangle$ to $|16\rangle$. \label{fig_Rb_ori}}
\end{figure}

Fig.~\ref{fig_Rb_ori} depicts all the couplings including the intended ones (thick blue arrows) and the unintended ones (thin red arrows)  from the lasers and cavity modes in the $^{87}$Rb atomic system with the levels involved labeled as $\ket{1} \sim \ket{16}$. From this, the Hamiltonian of the whole system, in a suitable rotating frame, can be written as
\begin{equation}
	H= {\rm h} \left[\delta |1\rangle \langle 1|+  \Omega_1 \left(|2\rangle \langle 1|+\frac{1}{c_{2,1}}\sum_{i=6}^8 c_{i,1} |i\rangle \langle 1| + \mathrm{H.c.}\right) + H_{\text{rig}}+H_{\text{flt}}\right]
\end{equation}
where h is the Plank constant; $\delta$ is the detuning of the first laser from the transition between the $|1\rangle$ and $|2\rangle$ states; $\Omega_1$ is the Rabi frequency of the first laser driving; $c_{j,i}$ is the Clebsch-Gordan coefficient corresponding to the $|i\rangle - |j\rangle$ coupling which determines the relative coupling strengths of the couplings induced by the same laser or cavity coupling;
\begin{equation}\label{}
	H_{\text{rig}}=
	\Omega_2|3 \rangle \langle 2|+
	g_{\rm t}|4\rangle \langle 3|\hat{a}_{\rm t}^\dagger+
	g_{\rm o}|5 \rangle\langle 4|\hat{a}_{\rm o}^\dagger+ \text{H.c.} \;;
\end{equation}
and
\begin{equation}\label{}
	\begin{split}
		H_{\text{flt}}=&-\Delta_7 |7\rangle\langle 7|-\left( \Delta_7+\Delta_8\right) |8\rangle\langle 8|-
		\Delta_{10} |10\rangle\langle 10|-\left( \Delta_{10}+\Delta_{11}\right) |11\rangle\langle 11|-\\
		&\Delta_{13} |13\rangle\langle 13|+
		\Delta_{15} \left(|15\rangle\langle 15|+
		|16\rangle\langle 16|\right)+\\
		&\left\{ \frac{\Omega_2}{c_{3,2}} \sum_{j=9}^{11} \sum_{i=6}^8 c_{j,i} \left|j\rangle \langle i\right| +
		\frac{g_{\rm t}}{c_{4,3}} \left[ \sum_{i=9}^{11} \left(c_{i,4}\left|4\rangle\langle i\right|+c_{12,i}\left|12\rangle\langle i\right|+c_{13,i}\left|13\rangle\langle i\right|\right) \right]\hat{a}_{\rm t}^{\mathrm{f}\dagger} + \right.\\
		&\left. \frac{g_{\rm o}}{c_{5,4}} \left[ \sum_{i=12}^{13} \left(c_{5,i}\left|5\rangle\langle i\right|+c_{14,i}\left|14\rangle\langle i\right|+c_{15,i}\left|15\rangle\langle i\right|+c_{16,i}\left|16\rangle\langle i\right|\right) \right]\hat{a}_{\rm t}^{\mathrm{f}\dagger} +\mathrm{H.c.} \right\},
	\end{split}
\end{equation}
where $\Delta_i$ are the Rb atomic energy splittings as shown in Fig. \ref{fig_Rb_ori}, and $\Omega_2$, $g_{\rm t}$, and $g_{\rm o}$ are the Rabi frequencies of the second laser driving, telecom-photon cavity coupling and visible-photon cavity coupling, respectively, and $\hat{a}_{\rm t}^\dagger$, $\hat{a}_{\rm o}^\dagger$ are the creation operators of the right telecom and visible photon, respectively. The creation operators for faulty ones are $\hat{a}_{\rm t}^{\mathrm{f}\dagger}$ and $\hat{a}_{\rm o}^{\mathrm{f}\dagger}$ assuming all the telecom or visible faulty photons are the same.

The atomic decay and the cavity leakage are characterized by the Lindblad operators below:

\begin{align}
	L_1=\sqrt{\gamma_{2}} |1\rangle\langle 2|,\quad L_2=\sqrt{\gamma_{3a}} |2\rangle\langle 3|,\quad L_3=\sqrt{\gamma_{3b}} |4\rangle\langle 3|,\quad 
	L_4=\sqrt{\gamma_{4}} |5\rangle\langle 4|,\quad \\
	L_5=\sqrt{\kappa_{\rm t}}\hat{a}_{\rm t},\quad 
	L_6=\sqrt{\kappa_{\rm o}}\hat{a}_{\rm o},\quad 
	L_7=\sqrt{\kappa_{\rm t}}\hat{a}_{\rm t}^{\mathrm{f}},\quad 
	L_8=\sqrt{\kappa_{\rm o}}\hat{a}_{\rm o}^{\mathrm{f}}.\quad 
\end{align}
where $\gamma_i$ is the atomic decay rate of the $|i\rangle$ level, especially, $\gamma_{3a}$ is the decay rate from the $|3\rangle$ state to the $|2\rangle$ state and $\gamma_{3b}$ to the $|4\rangle$ state. The decay rates $\kappa_{\rm t}$ and $\kappa_{\rm o}$ characterize the rate of the telecom- and visible-photon cavity photons going into the fibers, respectively. The cavity loss is not considered here because it is integrated into the fiber loss and/or considered in the entanglement swap section. The decay from the faulty levels has been neglected because the population in the faulty manifold is small. 

Regarding the parameters used in the simulation, the Rb atom has energy differences $\Delta_7=157$MHz, $\Delta_8=267$MHz, $\Delta_{10}=50.2$MHz, $\Delta_{11}=75.3$MHz, $\Delta_{13}=817$MHz, $\Delta_{15}=6.83$GHz and linewidths $\gamma_{2}=2\pi\times6.1$MHz, $\gamma_{3a}=2\pi\times0.19$MHz, $\gamma_{3b}=2\pi\times1.5$MHz, $\gamma_{4}=2\pi\times5.7$MHz. Besides, from the cavity simulation, we get the intended light proportions to be $p_1=0.98$, $p_2=0.99$, $p_3=0.83$, and $p_4=0.67$, and the cavity loss rates to be $\kappa_{\rm t}=2\pi\times1.5$GHz and $\kappa_{\rm o}=2\pi\times1.0$GHz. Besides, we also get the cooperativities for the telecom- and visible-photon cavities $C_{\rm t}=\frac{g_{\rm t}^2}{\kappa_{\rm t}\left(\gamma_{3a}+\gamma_{3b}\right)}=34.4\pm5.0$ and $C_{\rm o}=\frac{g_{\rm o}^2}{\kappa_{\rm o}\gamma_{4}}=11.2\pm2.2$ and those determine $g_{\rm t}$ and $g_{\rm o}$. Moreover, based on the Rb emitter simulation, the optimized $\Omega_2=2.0$GHz, $\delta=1.73$GHz, and $\Omega_1$ is at the order of 100 MHz and time-dependent with the temporal profile shown in Fig. 2(d) in the main text.

With the parameters above, we numerically calculate the system evolution based on the master equation
\begin{equation}
	\frac{\mathrm{d}\rho}{\mathrm{d}t}=\frac{\text{i}}{\rm h}[H,\:\rho]+
	\sum_{k}\left(\hat{L}_{k}\rho\hat{L}_{k}^{\dagger}-
	\frac{1}{2}\left\{ \hat{L}_{k}^{\dagger}\hat{L}_{k},\:\rho\right\} \right)\;.
	\label{MaOri}
\end{equation}
and get the temporal profiles of the emitted photons as shown in Fig. 2(d) in the main text. Based on the temporal profiles, we model the entanglement generation protocol with two Rb emitters. We sample the situations with varied cavity coupling and different moments within each time bin when the photons hit the photon detectors. As a result, leaving the fiber loss during distant transmission and inefficient detection aside, we get a success rate of 0.49 for generating the entangled visible photon pairs with an entangled state
\begin{equation}
	\rho_0=\left(\begin{array}{cccc}
		0 & 0 & 0 & 0\\
		0 & 0.5 & 0.48 & 0\\
		0 & 0.48 & 0.5 & 0\\
		0 & 0 & 0 & 0
	\end{array}\right).
	\label{}
\end{equation}
The codes of this Rb-cavity system simulation and elementary entanglement generation can be found in ref~\cite{codeCollection}.

\section{Suppression of faulty-polarization errors} \label{App_Rb_leakage}
\setcounter{figure}{0}
\setcounter{equation}{0}

\begin{figure}
	\centering
	\captionsetup{justification=centering}
	\includegraphics[width = 0.96\textwidth]{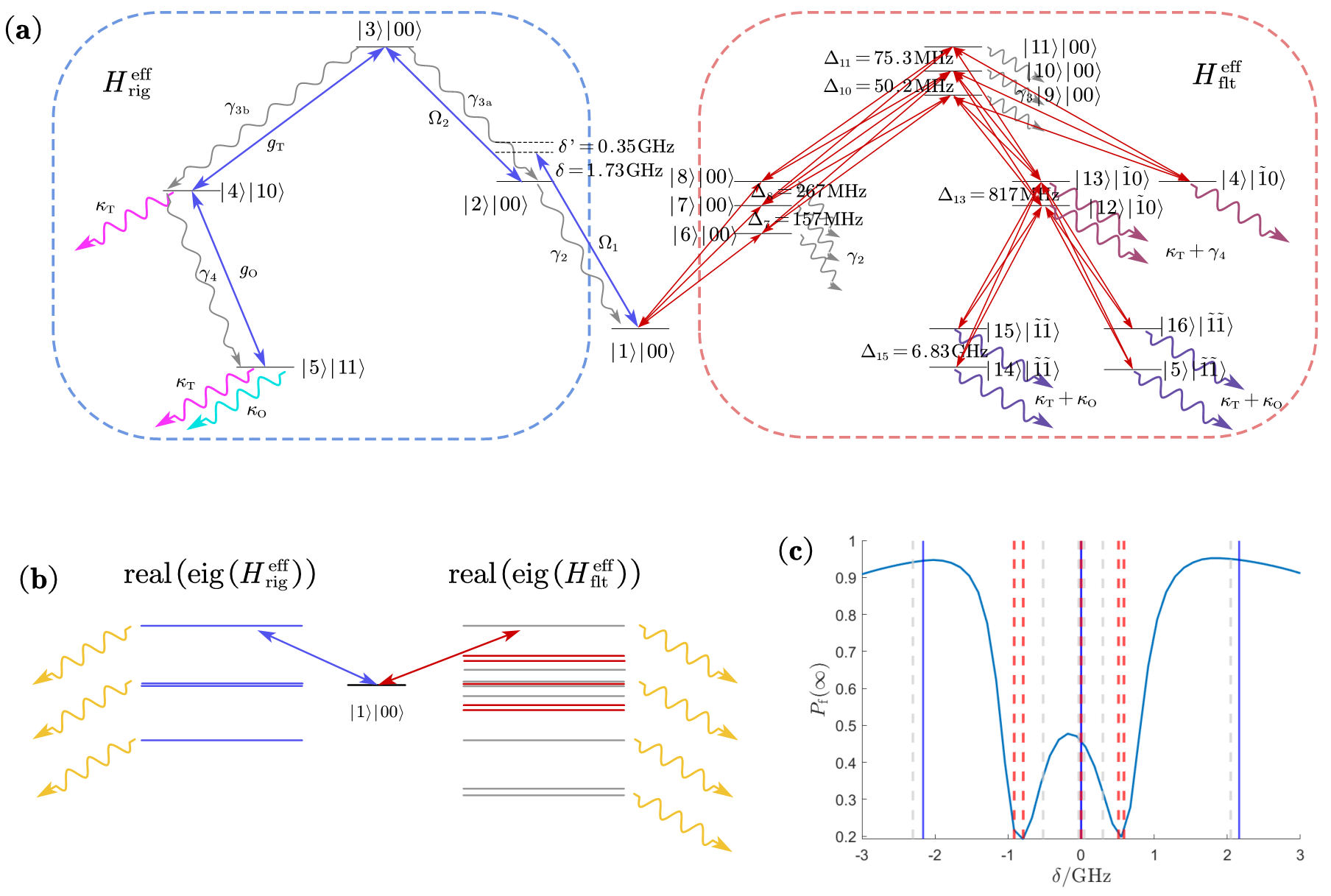}
	\caption{(a)Energy level diagram of the Rb-cavity system. Each quantum state is a direct product of the atomic state (first ket) and the photonic state (second ket). The atomic states are named with numbers in the decimal system. The two digits in the ket of the photonic state are either 0 or 1, meaning the vacuum or one-photon states in the telecom- (first digit) and visible- (second digit) photon cavities. The $\tilde{1}$ with a tilde means the one-photon state of faulty photons. The diagram is arranged in the way that the energy levels in the left dashed blue rounded rectangle belong to the intended subsystem and the ones in the right dashed red round rectangle consist of the faulty subsystem. (b) The spectrum, or the real parts of the eigenvalues of the effective Hamiltonians of the intended (left) and faulty (right) subsystems, respectively. The gray lines in the faulty subsystem have neglectful coupling to the $|1\rangle|00\rangle$ state. The double arrows point to the targeted states used in our simulations. (c) Right photon emission proportion over the sweep of the first laser detuning $\delta$. Vertical lines denote the eigenenergies of the right subsystem (solid blue) and the faulty subsystem (dashed red and gray). The code of this calculation can be found in ref~\cite{codeCollection}. \label{fig_Rb_rn}}
\end{figure}

For better analysis of the dynamic of the whole Rb-cavity system, we derive the energy-level diagram of the system incorporating the photonic states shown in Fig \ref{fig_Rb_rn}(a) where we have separated the intended path and the faulty paths into two sides. Since the first laser driving $\Omega_1$ is about one order weaker than the driving from the second laser and cavity couplings, it is suitable to first neglect the couplings induced by it and separate the excited systems into two parts: Specifically, the intended subsystem driven by the $\sigma_+$ polarized component of the first laser light and the faulty subsystem driven by the $\sigma_-$ polarized component, as circled in with blue and red dash lines in Fig. \ref{fig_Rb_rn}(a). These two subsystems are governed by the two Hamiltonians $H_{\text{rig}}$ and $H_{\text{flt}}$, respectively, which are defined as
\begin{equation}
	H_{\text{rig}}^{\text{eff}}=H_{\text{rig}}+\frac{h}{2\mathrm{i}}\sum_{k=1}^{6}L_k^\dagger L_k
\end{equation}
and 
\begin{equation}
	H_{\text{flt}}^{\text{eff}}=H_{\text{flt}}+\frac{h}{2\mathrm{i}}\left(\sum_{k=7}^{8}L_k^\dagger L_k + L_4^\dagger L_4\right)\;,
\end{equation}

Diagonalizing these two effective Hamiltonians and taking the real parts of the eigenvalues, one can get the spectrum of two subsystems as shown in Fig. \ref{fig_Rb_rn}(b), while the imaginary parts of the eigenvalues correspond to the effective decay rates of the eigenstates or the dressed states. From the spectrum, we find it possible to couple strongly to specific dressed states of the right subspace while weakly coupling to the ones from the faulty subsystem. Thus, having the right detuning of the first laser can strongly suppress any driving from the $\sigma_-$ component. Fig. \ref{fig_Rb_rn}(c) shows the right photon emission rate sweeping over the detuning $\delta$. One can see that the right photon emission rate peaks at regions where one of the dressed states of the right subsystem is resonantly driven while the coupling to the dressed states of the faulty subsystem is weak.

\section{Stabilization of Population in time-bins}
\setcounter{figure}{0}
\setcounter{equation}{0}

\begin{figure}
	\centering
	\captionsetup{justification=centering}
	\includegraphics[width = 0.65\textwidth]{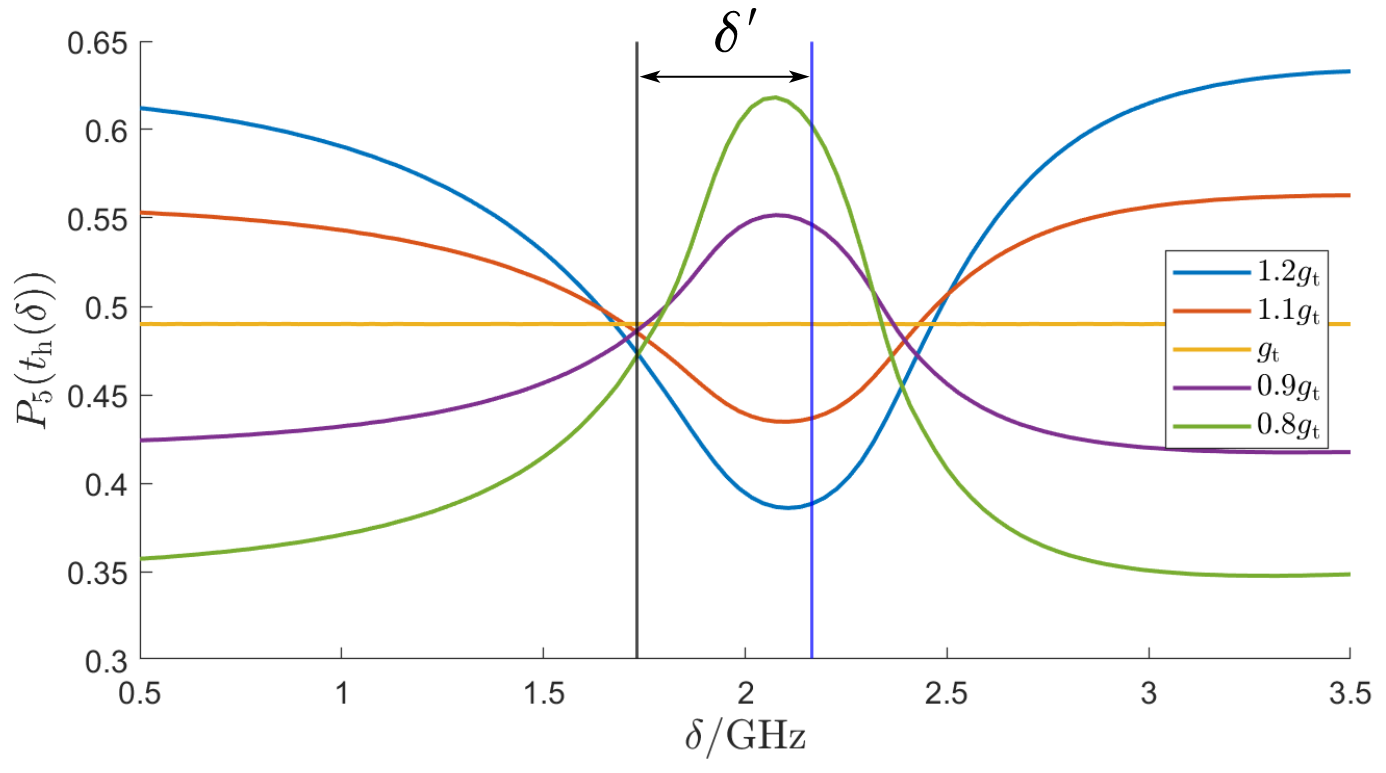}
	\caption{The fluctuation of the possibility that photons are emitted in the early time bin caused by the fluctuation in the atom-telecom-cavity coupling. The $t_{\rm h}(\delta)$ is the calibrated time for each $\delta$ making sure that, when the telecom-cavity coupling $g_{\rm t}$ is at its averaged value, the photon emission possibility in the early time-bin is always 0.49 (the yellow horizontal line). Different lines show different population emission possibilities in the early time bin corresponding to different telecom-cavity coupling strengths. The blue vertical line marks the eigenenergy of the targeted state of the intended subsystem. The black vertical line marks the sweet spot where the emission possibilities are efficiently stabilized. The code of this calculation can be found in ref~\cite{codeCollection}. \label{fig_stable} }
\end{figure}

Our simulation suggests that the Rb-telecom-cavity coupling has an averaged cooperativity of 34 with a standard deviation of 5.0. This results in roughly 10\% standard deviation in the Rb-telecom-cavity coupling strength $g_{\rm t}$, which is the main cause of the fluctuation in the time-bin population of the photons. However, by sweeping the first laser frequency, we find sweet spots where this fluctuation is efficiently suppressed as shown in Fig.~\ref{fig_stable}.

These sweet spots can be understood from the dressed states of the subspace spanned with $\ket{2}\ket{00}$, $\ket{3}\ket{00}$, $\ket{4}\ket{10}$ states. Coupled by the second laser and the telecom cavity field, These three states form a typical $\Lambda$ system whose eigenstates are close to the ones in our system. The corresponding eigenstate to the one we are targeting reads
\begin{equation}
	\ket{\psi_{\rm tar}}=\frac{1}{\sqrt{2}}\left( \ket{3}\ket{00} + \frac{g_{\rm t}}{\sqrt{g_{\rm t}^2+\Omega_2^2}}\ket{4}\ket{10}+
	\frac{\Omega_2}{\sqrt{g_{\rm t}^2+\Omega_2^2}}\ket{2}\ket{00} \right).
\end{equation}
In this picture, one can view the atomic driving procedure as that the population is driven from the $\ket{1}\ket{00}$ state to the $\ket{\psi_{\rm tar}}$ with the first laser and then the $\ket{\psi_{\rm tar}}$ experiences cavity decay leaking the telecom photon into the fiber. In this case, the effective first laser driving becomes:
\begin{equation}
	\Omega_1^{\rm eff}
	=\frac{\Omega_2\Omega_1}
	{\sqrt{2}\sqrt{g_{\rm t}^2+\Omega_2^2}},
\end{equation}
and the effective cavity decay rate turns into:
\begin{equation}
	\kappa_{\rm t}^{\rm eff}
	=\frac{g_{\rm t}^2 \kappa_t}
	{2\left(g_{\rm t}^2+\Omega_2^2\right)}.
\end{equation}
Therefore, the fluctuating $g_{\rm t}$ affects positively on $\kappa_{\rm t}^{\rm eff}$ but negatively on $\Omega_1^{\rm eff}$. By introducing a detuning $\delta'$ respecting to the $\ket{1}\ket{00}-\ket{\psi_{\rm tar}}$ transition, one can tune the relative weights of these two effects and let them maximally compensate each other. As a result, the transition rate is stabilized over the fluctuation of $g_{\rm t}$.

\section{Trapping and cavity designs}

To couple an atom with two cavity fields, we design two nanophotonic cavities at the target frequencies fabricated parallel to each other, combined with a trap geometry that enables trapping an atom between the cavities. The individual cavities are designed by quadratic tapering of the filling fractions on a nanobeam waveguide\cite{Quan2011}. For practical fabrication considerations, both cavities are designed with the same thickness. Due to its larger wavelength, the TE cavity at 1470 nm, requires a larger thickness for mode confinement and subsequently a high quality factor. However, for TE mode cavities for 795 nm, these thicknesses result in low modal overlap in the atom trapping region due to large mode confinement. To overcome this issue, we make use of the TM mode cavity design for 795 nm cavity, which requires a larger thickness while maintaining decent modal overlap in the trapping region.

Quality factors of both the 780 nm and 1470 nm cavities are individually optimized and brought close to each other to form a stable trap geometry. Atoms can be trapped near similar nanophotonic devices by bringing the tweezer adiabatically on top of these devices and trapping the atom in lattices formed by the incident and reflected tweezer\cite{Thompson2013, Menon2023, Goban2014}. In the case of two parallel cavities, trapping on top of any one of the two cavities results in minimal coupling to the other cavity due to minimal cavity mode at the trapping region. To have a significant coupling strength to both cavities, we propose trapping in between the two devices as an alternate implementation. If the separation between the parallel cavities is larger than the wavelength, and the diffraction-limited spot size of the trapping tweezer, atoms can be trapped in the normal tweezer. However, the large rayleigh range of a normal tweezer implies an atomic wave function larger than the device's thickness. This results in a large variation in the cavity coupling strength experienced by the atom from shot to shot. To overcome this, we restrict the separation between the two parallel cavities to be significantly less than the trapping tweezer wavelength. In this case, the incident tweezer beam is reflected by the combined device structure to form a lattice-like potential similar to a single device case. For a reasonable trapping potential of atoms, the separation between the devices should be smaller than half the trapping tweezer wavelength.

While the cavities were designed to have high quality factors individually, when they are kept next to each other, the mode from one cavity can leak into the other. These additional extrinsic losses $\kappa_{\rm ext}$ through the second waveguide result in lower overall quality factors ($Q= \frac{\omega}{\kappa_{\rm ext} + \kappa_{\rm int}}$; $\omega$ is the resonant frequency; $\kappa_{\rm int}$ is the intrinsic losses not coupled into any waveguides; $\kappa_{\rm ext}$ includes losses into the intended waveguide mode and the additional losses to the nearby waveguide). This loss is a function of separation between the two cavities, where the loss increases with reduced separation between the cavities. 

We iterate between device separation, thickness, and refractive index of the material to find a deep trap potential while maintaining high quality factors required for the results presented in the paper. To iterate over the refractive index, we assume tuning of the silicon enrichment ratio in the silicon nitride. However, the ratio of silicon in the silicon nitride also modifies the bandgap of the material. As the nitride content is reduced, the refractive index and the bandgap approaches that of bare silicon\cite{Philipp1973}. To avoid the above bandgap excitation using 780 nm laser involved in the protocol, we assume the refractive index of 2.6, assumed in the paper, results in a bandgap larger than 1.6 eV.

Atoms can be trapped in regions where cavity fields have predominantly linear polarization direction on top of nanophotonic devices \cite{Menon2020}. However, moving the trap to the edge of the devices results in cavity field polarization to have components in more than one dimension in the trapping region, without specific control over the phase between the polarizations. This results in varying proportions of contributions from $\sigma_+$, $\sigma_-$, and $\pi$ in the trapping region. The purity of the intended sigma polarization is marked in Fig. \ref{fig_Rb_ori}. To overcome this issue, we make a careful selection of the states involved in the scheme. The level scheme involved in the protocol is chosen in a way that under correct excitation, only $\sigma_-$  polarization of the cavity fields couple between atomic states. The corrections coming from the faulty excitation are labeled in the figure and are accounted for.

\section{Optimized Number of Repeaters}
\setcounter{figure}{0}
\setcounter{equation}{0}

\begin{figure}
	\centering
	\captionsetup{justification=centering}
	\includegraphics[width = 0.98\textwidth]{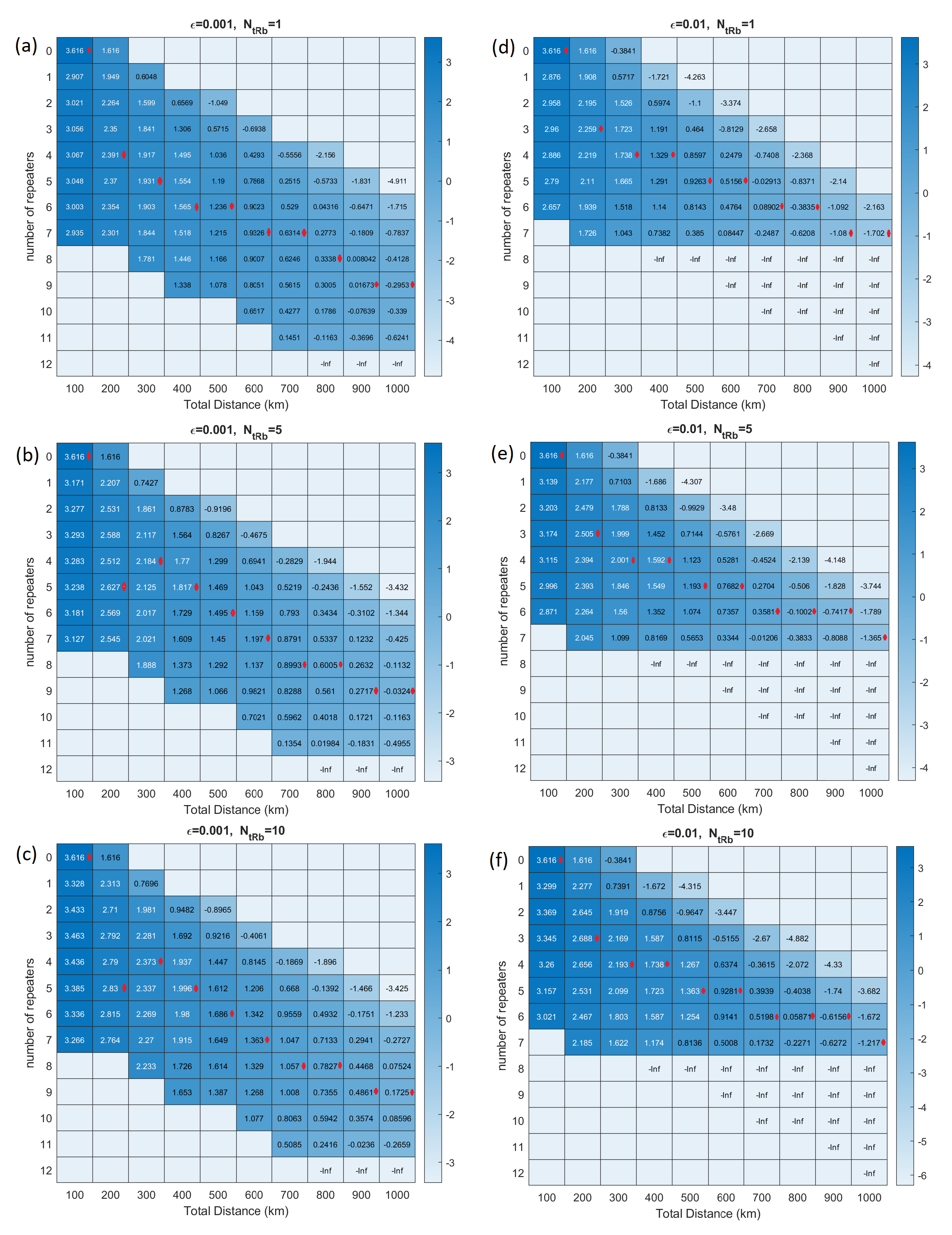}
	\caption{Scanning of the secret key rate per segment $R_{\rm SK}/N_{\rm seg}$ over the number of repeaters and total distance for different swapping errors $\epsilon$ and number of Rb atoms $N_{\rm tRb}$ for entanglement transfer. The value in the heatmaps are $\log_{10}\left(R_{\rm SK}/N_{\rm seg}\right)$, hence "-inf" meaning $-\infty$ means $R_{\rm SK}=0$ and the empty entries are not calculated. The red dots in the heatmaps identify the maximum secret key rate per segment for each total distance determining the optimum number of repeaters. \label{fig_n_rep_sc}}
\end{figure}

The scanning of the secret key rate per segment $R_{\rm SK}/N_{\rm seg}$ over the number of repeaters and total distance is shown in Fig.~\ref{fig_n_rep_sc}.


\end{document}